\documentclass[aip,reprint,numeric]{revtex4-1}
\usepackage{amsfonts,amssymb,amsmath,graphicx,enumerate}
\usepackage{amsthm}
\usepackage{multirow}
\usepackage{color}
\usepackage{hyperref}

\begin{document}

\title{Unified functional network and nonlinear time series analysis for complex systems science: The \texttt{pyunicorn} package}

\author{Jonathan F. Donges}
    \email{donges@pik-potsdam.de}
    \affiliation{Potsdam Institute for Climate Impact Research, P.O. Box 601203, D-14412 Potsdam, Germany}
    \affiliation{Stockholm Resilience Centre, Stockholm University, Kr\"aftriket 2B, 114 19 Stockholm, Sweden}
\author{Jobst Heitzig}
    \affiliation{Potsdam Institute for Climate Impact Research, P.O. Box 601203, D-14412 Potsdam, Germany}
\author{Boyan Beronov}
    \affiliation{Potsdam Institute for Climate Impact Research, P.O. Box 601203, D-14412 Potsdam, Germany}
\author{Marc Wiedermann}
    \affiliation{Potsdam Institute for Climate Impact Research, P.O. Box 601203, D-14412 Potsdam, Germany}
    \affiliation{Department of Physics, Humboldt University Berlin, Newtonstr.~15, D-12489 Berlin, Germany}
\author{Jakob Runge}
    \affiliation{Potsdam Institute for Climate Impact Research, P.O. Box 601203, D-14412 Potsdam, Germany}
    \affiliation{Department of Physics, Humboldt University Berlin, Newtonstr.~15, D-12489 Berlin, Germany}
\author{Qing Yi Feng}
    \affiliation{Institute for Marine and Atmospheric Research Utrecht (IMAU), Department of Physics and Astronomy, Utrecht University, Utrecht, The Netherlands}
\author{Liubov Tupikina}
    \affiliation{Potsdam Institute for Climate Impact Research, P.O. Box 601203, D-14412 Potsdam, Germany}
    \affiliation{Department of Physics, Humboldt University Berlin, Newtonstr.~15, D-12489 Berlin, Germany}
\author{Veronika Stolbova}
    \affiliation{Potsdam Institute for Climate Impact Research, P.O. Box 601203, D-14412 Potsdam, Germany}
    \affiliation{Department of Physics, Humboldt University Berlin, Newtonstr.~15, D-12489 Berlin, Germany}
\author{Reik V. Donner}
    \affiliation{Potsdam Institute for Climate Impact Research, P.O. Box 601203, D-14412 Potsdam, Germany}
\author{Norbert Marwan}
    \affiliation{Potsdam Institute for Climate Impact Research, P.O. Box 601203, D-14412 Potsdam, Germany}
\author{Henk A. Dijkstra}
    \affiliation{Institute for Marine and Atmospheric Research Utrecht (IMAU), Department of Physics and Astronomy, Utrecht University, Utrecht, The Netherlands}
\author{J\"urgen Kurths}
   \affiliation{Potsdam Institute for Climate Impact Research, P.O. Box 601203, D-14412 Potsdam, Germany}
   \affiliation{Department of Physics, Humboldt University Berlin, Newtonstr.~15, D-12489 Berlin, Germany}
   \affiliation{Institute for Complex Systems and Mathematical Biology, University of Aberdeen, Aberdeen AB24 3FX, United Kingdom}
   \affiliation{Department of Control Theory, Nizhny Novgorod State University, Gagarin Avenue 23, 606950 Nizhny Novgorod, Russia}

\date{\today}                                           

\begin{abstract}
We introduce the \texttt{pyunicorn} (Pythonic unified complex network and recurrence analysis toolbox) open source software package for applying and combining modern methods of data analysis and modeling from complex network theory and nonlinear time series analysis. \texttt{pyunicorn} is  a fully object-oriented and easily parallelizable package written in the language Python. It allows for the construction of functional networks such as climate networks in climatology or functional brain networks in neuroscience representing the structure of statistical interrelationships in large data sets of time series and, subsequently, investigating this structure using advanced methods of complex network theory such as measures and models for spatial networks, networks of interacting networks, node-weighted statistics or network surrogates. Additionally, \texttt{pyunicorn} provides insights into the nonlinear dynamics of complex systems as recorded in uni- and multivariate time series from a non-traditional perspective by means of recurrence quantification analysis (RQA), recurrence networks, visibility graphs and construction of surrogate time series. The range of possible applications of the library is outlined, drawing on several examples mainly from the field of climatology.
\end{abstract}

\maketitle

\begin{quotation}
Network theory and nonlinear time series analysis provide powerful tools for the study of complex systems in various disciplines such as climatology, neuroscience, social science, infrastructure or economics. In the last years, combining both frameworks has yielded a wealth of new approaches for understanding and modeling the structure and dynamics of such systems based on the statistical analysis of network or uni- and multivariate time series. The \texttt{pyunicorn} software package (available at \texttt{https://github.com/pik-copan/pyunicorn}) facilitates the innovative synthesis of methods from network theory and nonlinear time series analysis in order to develop novel integrated methodologies. This paper provides an overview of the functionality provided by \texttt{pyunicorn}, introduces the theoretical concepts behind it and provides examples in the form of selected use cases mainly in the fields of climatology and paleoclimatology.
\end{quotation}

\section{Introduction}
\label{sec:intro}

Complex network theory~\citep{Albert2002,Newman2003,Boccaletti2006,Cohenbook2010,Newmanbook2010} and nonlinear time series analysis~\citep{Abarbanel1996,Sprott2003chaos,Kantz2004} provide two complementary perspectives on the structure and dynamics of complex systems. Historically, the investigation of complex networks has focussed on the structure of interactions (links or edges) between the possibly large number of subsystems (nodes or vertices) of a complex system, e.g. searching for universal properties like scaling behavior or identifying specific classes of nodes such as bottlenecks that are particularly important transmitters for flows on the network. In contrast, nonlinear time series analysis emphasized dynamical aspects such as predictability, chaos, dynamical transitions or bifurcations in the observed or modeled time-dependent state variables of complex systems. For a long time, these communities were mostly disconnected and, particularly, applied distinct software tools such as \texttt{igraph}~\citep{Csardi2006} or \texttt{networkx}~\citep{Schult2008exploring} for analyzing complex networks and the classical \texttt{TISEAN} package for nonlinear time series analysis~\citep{Hegger1999}.

In the last several years, two strands of research have been taken advantage of the synergies obtained by combining complex network theory and nonlinear time series analysis. On the one hand, the analysis of functional networks put forward in neuroscience~\citep{Zhou2006,zhou2007structure,Bullmore2009} and climatology~\citep{Tsonis2004,Tsonis2008a,Yamasaki2008,Donges2009a,Donges2009b,Donges2015} as well as other application areas such as economics and finance~\citep{huang2009network}, applies methods from linear and nonlinear time series analysis to construct networks of statistical interrelationships among a set of time series and, subsequently, studies the resulting functional networks by means of methods from complex network theory. On the other hand, network-based time series analysis investigates the dynamical properties of complex systems' states based on uni- or multivariate time series data using methods from network theory~\citep{Donner2011a}. Various types of time series networks have been proposed for performing this type of analysis, including recurrence networks based on the recurrence properties of phase space trajectories~\citep{Xu2008,Marwan2009,Donner2010b,Donges2012}, transition networks encoding transition probabilities between different phase space regions~\citep{Nicolis2005} and visibility graphs representing visibility relationships between data points in a time series~\citep{Lacasa2008,Donner2012visibility,Donges2013}.

The purpose of this paper is to introduce the Python software package \texttt{pyunicorn}, which implements methods from both complex network theory and nonlinear time series analysis, and unites these approaches in a performant, modular and flexible way. Thereby, \texttt{pyunicorn} allows to easily apply recently developed techniques combining these perspectives, such as functional networks and network-based time series analysis. Furthermore, the software allows to conveniently generate new syntheses of existing concepts and methods from both fields that can lead to novel methodological developments and fruitful applications in the future. While in this tutorial paper, the work flow of using \texttt{pyunicorn} is mainly illustrated drawing upon examples from climatology, the package is applicable to all fields of study where the analysis of (big) time series data is of interest, e.g. neuroscience~\citep{Bullmore2009,Subramaniyam2014,Subramaniyam2015}. In this paper, while we aim to give a practical overview on the functionality and possibilities of \texttt{pyunicorn}, we cannot provide a comprehensive reference or handbook due to space constraints. For such a reference, see the \texttt{pyunicorn} API documentation~\citep{Supplement} (see \texttt{pyunicorn} website for newest version).

This article is structured as follows: After a general introduction of \texttt{pyunicorn} and a discussion of the philosophy behind its implementation, software structure and related computational issues (Sect.~\ref{sec:intro}), \texttt{pyunicorn}'s capabilities for analyzing and modeling complex networks are described including general networks, spatial networks, networks of interacting networks or multiplex networks and node-weighted networks (Sec.~\ref{sec:networks}). Building on this, Sect.~\ref{sec:functional_networks} presents methods for constructing and analyzing functional networks from fields of multiple time series, including use cases demonstrating the application of climate network and coupled climate network analysis. Section \ref{sec:time_series} describes \texttt{pyunicorn}'s methods for performing nonlinear time series analysis using recurrence plots, recurrence networks and visibility graphs. Methods for generating surrogate time series, which are useful for both functional network and network-based time series analysis, are introduced in Sect.~\ref{sec:surrogates}. Finally, conclusions are drawn and some perspectives for future further development of \texttt{pyunicorn} are outlined (Sect.~\ref{sec:conclusions}).

\subsection{Implementation philosophy}
\label{sec:philosophy}

\begin{figure*}[tbh]
\begin{center}
\includegraphics[width=0.66 \textwidth]{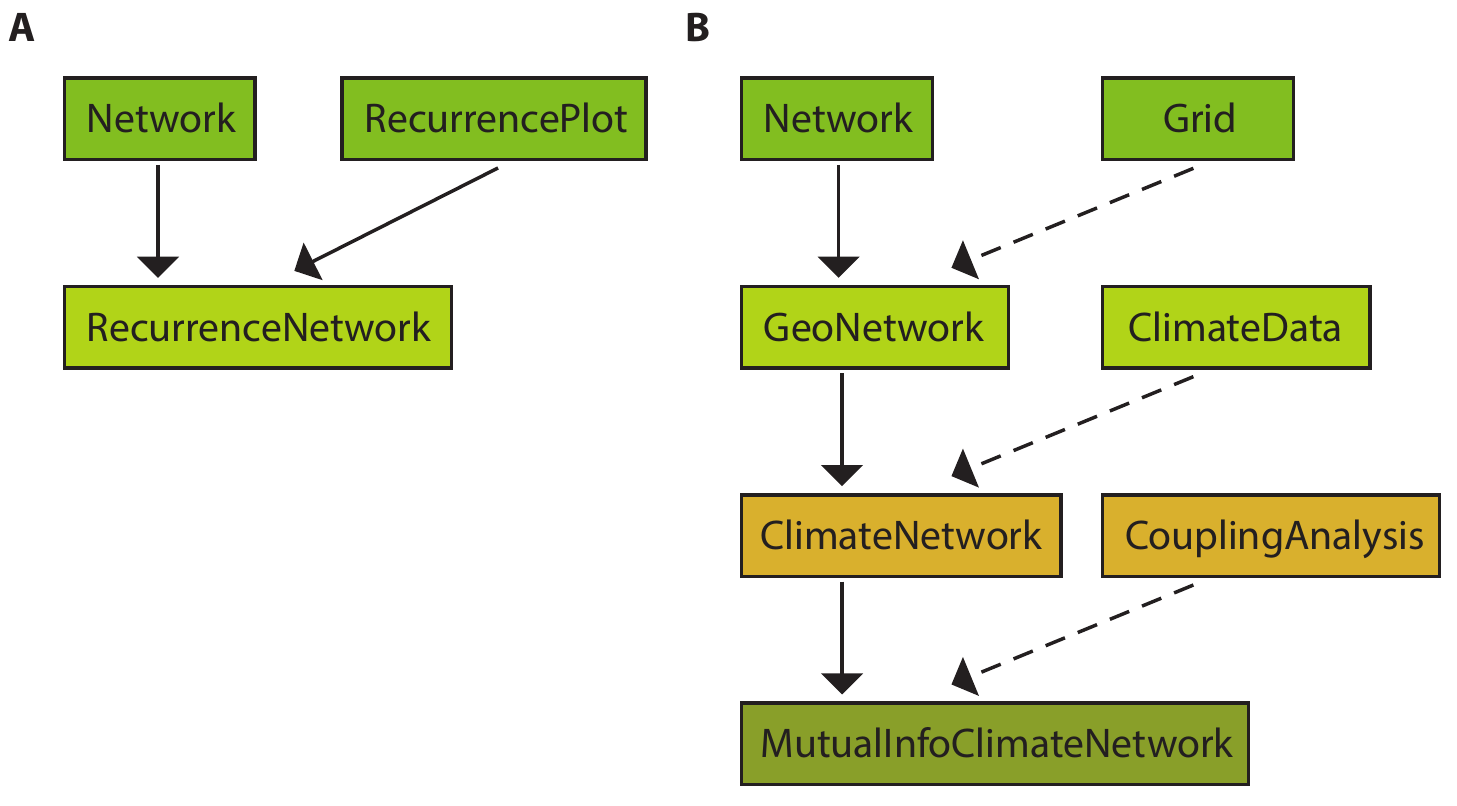}
\end{center}
\caption[]{
Examples for the software architecture of \texttt{pyunicorn} displayed as a Unified Modeling Language (UML) diagram of class relationships: ancestry of the (A) \texttt{timeseries.RecurrenceNetwork} (Sect.~\ref{sec:recurrence_networks}) and (B) \texttt{climate.MutualInfoClimateNetwork} classes (Sect.~\ref{sec:climate_networks}). Inheritance (class \texttt{B} inherits from class \texttt{A}, solid arrows) and object composition relationships (class \texttt{B} contains class \texttt{A}, dashed arrows) are indicated.}
\label{fig:uml}
\end{figure*}


\texttt{pyunicorn} is intended as an integrated container for a host of conceptionally related 
methods which have been developed and applied by the involved research groups for many years. 
Its aim is to establish a shared infrastructure for scientific data analysis by 
means of complex networks and non-linear time series analysis and it has 
greatly taken advantage from the backflow 
contributed by users all over the world. The code base has been fully open 
sourced under the BSD 3-Clause license.

With a focus on complex network methods, this software is a valuable complement 
to traditional non-linear time series analysis tools such as \texttt{TISEAN}~\citep{Hegger1999}. Its main mode of 
operation is to import, generate and export complex networks from time series 
or fields thereof, and to compute appropriate measures on these networks in 
order to derive insights into the causal structure and dynamical regimes of
underlying processes. While \texttt{pyunicorn}'s development has mostly 
accompanied advances in climatology and paleoclimatology, the generality of the 
network approach and its implementation of extensions to standard complex networks like 
spatio-temporal networks, node weighted measures, coupled 
functional networks and recurrence networks render the software widely 
applicable in numerous fields, e.g. medicine, neuroscience, sociology, 
economics and finance. Great care has been taken in linking to relevant 
publications from the method descriptions contained in the code and API documentation~\citep{Supplement}.

As the name suggests, the language chosen for the implementation is Python, 
which is very well established in scientific 
computing~\citep{Oliphant2007,Millman2011}. Due to the multiplicity of useful 
combinations of methods, 
there are no executables in \texttt{pyunicorn}, but the library 
is intended to be used by small Python scripts. Its object-oriented software 
architecture allows for clean and flexible code representing the logical 
interrelationships and dependencies between the various concepts and methods 
(Sect.~\ref{sec:software_structure}). For example, the class 
\texttt{RecurrenceNetwork} (Sect.~\ref{sec:recurrence_networks}) inherits from 
both the \texttt{Network} (Sect.~\ref{sec:networks_basics}) and 
\texttt{RecurrencePlot} classes (Sect.~\ref{sec:rqa}), thus reflecting the 
mathematical definition and historical development of recurrence network 
analysis (Fig.~\ref{fig:uml}A). Following a similar reasoning behind the 
implementation of a class hierarchy, the climate network class 
\texttt{MutualInfoClimateNetwork} (Sect.~\ref{sec:climate_networks}) inherits 
from the \texttt{Network} class via the intermediate parent classes 
\texttt{GeoNetwork} (Sect.~\ref{sec:spatial_networks}) and 
\texttt{ClimateNetwork} (Sect.~\ref{sec:climate_networks}), additionally 
including several object composition relationships on the way 
(Fig.~\ref{fig:uml}B).

While ensuring accessibility and maintainability among scientists in the disciplines mentioned above, this design 
facilitates fully flexible use of the package, from interactive local sessions 
in \texttt{IPython}~\citep{Perez2007} to massively parallel computations on cluster 
architectures. For several years now, \texttt{pyunicorn} has been successfully 
deployed on Linux, Mac OS X and Windows systems as well as UNIX high 
performance clusters.

Besides \texttt{Numpy}~\citep{Walt2011} and 
\texttt{Scipy}~\citep{Jones2001scipy}, 
which are among the most widely spread libraries for scientific computing in Python, 
\texttt{pyunicorn}'s only hard dependency is the \texttt{igraph} network 
analysis package~\citep{Csardi2006}. \texttt{pyunicorn} does not possess its own graphical 
interface, but where visual output is meaningful, helper methods exist for 
plotting with \texttt{matplotlib}~\citep{Hunter2007}, which is especially convenient in 
\texttt{IPython}. Interfaces to tools for advanced network visualization focussing on spatial networks~\citep{Nocke2015} 
such as \texttt{CGV}~\citep{tominski2009cgv,tominski2011information} are 
provided via \texttt{pyunicorn}'s input-output capabilities. 
Commented examples for typical use cases are provided by the extensive software documentation.

\subsection{Software structure}
\label{sec:software_structure}

\begin{table*}[bth]
\caption{Structure of the \texttt{pyunicorn} software package listing the most important classes belonging to each submodule (selection for brevity).}
\begin{center}
\begin{tabular}{lllll}
\hline
\textbf{core} & \textbf{funcnet} & \textbf{climate} & \textbf{timeseries} & \textbf{utils} \\
\hline
(Sect.~\ref{sec:networks}) & (Sect.~\ref{sec:functional_networks}) & (Sects.~\ref{sec:climate_networks}, \ref{sec:coupled_climate_nets}) & (Sects.~\ref{sec:time_series}, \ref{sec:surrogates}) & (Sect.~\ref{sec:intro}) \\
\texttt{Network} & \texttt{CouplingAnalysis} & \texttt{ClimateNetwork} & \texttt{RecurrencePlot} & \texttt{mpi} \\
\texttt{GeoNetwork} & \texttt{CouplingAnalysisPurePython} & \texttt{CoupledClimateNetwork} & \texttt{CrossRecurrencePlot} & \texttt{navigator} \\
\texttt{InteractingNetworks} & & \texttt{TsonisClimateNetwork} & \texttt{JointRecurrencePlot} & \\
\texttt{ResNetwork} & & \texttt{SpearmanClimateNetwork} & \texttt{RecurrenceNetwork} & \\
\texttt{Data} & & \texttt{MutualInfoClimateNetwork} & \texttt{InterSystemRecurrenceNetwork} & \\
\texttt{Grid} & & \texttt{ClimateData} & \texttt{JointRecurrenceNetwork} & \\
 & & & \texttt{VisibilityGraph} & \\
 & & & \texttt{Surrogates} & \\
\hline
\end{tabular}
\end{center}
\label{tab:pyunicorn_structure}
\end{table*}%

The \texttt{pyunicorn} library is fully object-oriented and its inheritance and composition hierarchy reflects the relationships between the analysis methods in use (Fig.~\ref{fig:uml}). It consists of five subpackages (Tab.~\ref{tab:pyunicorn_structure}):

\begin{description}

\item[core] This name space contains the basic building blocks for general 
network analysis and modeling and is accessible after calling \texttt{import 
pyunicorn} (Sect.~\ref{sec:networks}). The backbone \texttt{Network} class provides numerous standard and 
advanced complex network statistics, measures and generative models as well as 
import and export capabilities to GraphML, GML, NCOL, LGL, DOT, DIMACS and other formats. 
\texttt{Grid} and \texttt{GeoNetwork} extend these functionalities with respect to 
spatio-temporally embedded networks, which can be imported from and exported to 
ASCII and NetCDF files via the \texttt{Data} class. \texttt{InteractingNetworks} 
provides advanced methods designed for networks of networks (or multiplex networks), while 
\texttt{ResNetwork} specializes in power grids transporting electric currents and related infrastructure networks.

\item[funcnet] Advanced tools for construction and analysis of general 
(non-climate) functional networks will be accommodated here. So far, 
\texttt{CouplingAnalysis} calculates cross-correlation, mutual information, 
mutual sorting information and their respective surrogates for large arrays of 
scalar time series (Sect.~\ref{sec:functional_networks}).

\item[climate] Building on top of \texttt{GeoNetwork} and \texttt{Data}, the 
\texttt{ClimateNetwork} class and its children facilitate the construction and 
analysis of functional networks representing the statistical interdependency 
structure within a field of time series, based on similarity measures such as 
lagged linear Pearson or Spearman correlation and mutual information 
(Sect.~\ref{sec:climate_networks}). \texttt{CoupledClimateNetwork} extends this capability to 
the study of interrelationships between two distinct fields of observables (Sect.~\ref{sec:coupled_climate_nets}).

\item[timeseries] This subpackage provides various tools for the analysis of 
non-linear dynamical systems and uni- and multivariate time series (Sect.~\ref{sec:time_series}). Apart from 
visibility graphs with time-directed measures (\texttt{VisibilityGraph} class), the focus lies on 
recurrence-based methods derived from the \texttt{RecurrencePlot} class. These 
include single, joint and cross-recurrence plots as well as standard, joint and 
inter-system recurrence networks, supporting time-delay embedding and 
several phase space metrics and are equipped with common measures of recurrence 
quantification analysis. \texttt{Surrogates} allow testing for the statistical significance of 
similarity measures by generating surrogate data sets under miscellaneous 
constraints from observable time series (Sect.~\ref{sec:surrogates}).

\item[utils] Currently this includes \texttt{MPI} parallelization support and 
an experimental interactive network navigator.

\end{description}

\subsection{General computational issues}
\label{sec:computation}


Most network measures are defined as aggregates of local information
obtained from topology, node weights and link attributes. Since
\texttt{pyunicorn} internally represents networks as sparse adjacency matrices,
it can handle large data sets. Until streaming algorithms are implemented for
all measures, the available amount of working memory (RAM) limits the size of 
networks which can be processed. Presently, many advanced methods are defined 
only for undirected networks. As sensible generalizations and applications come 
up, they are gradually incorporated into the code base.

\begin{figure}[bp]
\begin{center}
\includegraphics[width= 0.75 \columnwidth]{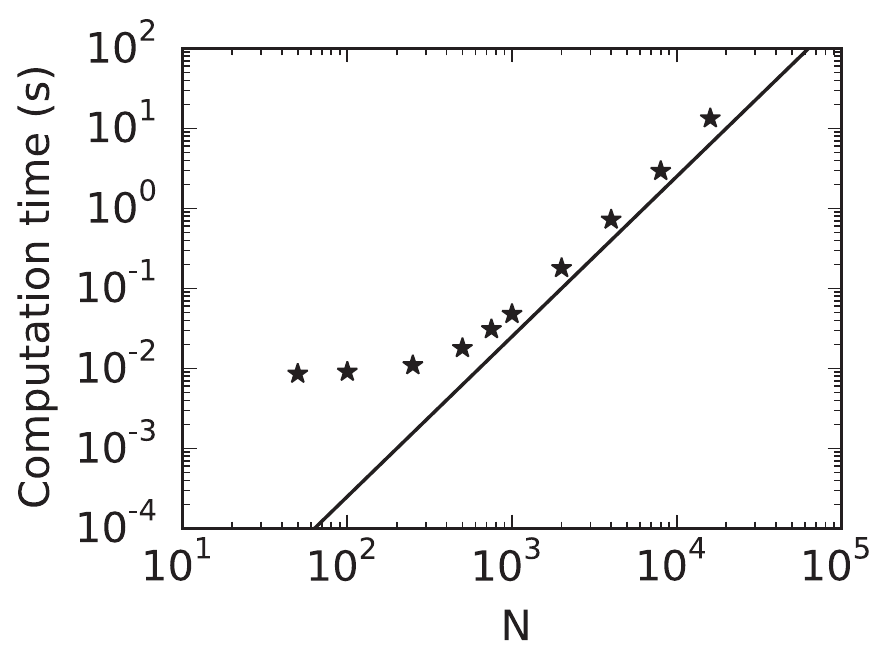}
\end{center}
\caption[]{Dependence of computation time for constructing recurrence networks (represented by \texttt{RecurrenceNetwork} objects) of varying size $N$ constructed from 
state vectors randomly drawn from the Lorenz attractor (see Fig.~\ref{fig:rn_lorenz}) on a dual-core Intel Core i5 CPU 
with 2.4 GHz running Mac OS X (star symbols). For larger $N$, computation time is proportional to $N^2$ (solid line) as expected
for the straightforward algorithm implemented for calculating the $N \times N$ recurrence matrix $\mathbf{R}$. Note that fast neighbor search algorithms, 
currently not implemented in \texttt{pyunicorn}, can reduce the number of required computations to $\mathcal{O}(N \log N)$ or even $\mathcal{O}(N)$ under certain conditions~\citep{Hegger1999}.
}
\label{fig:rn_stats}
\end{figure}

As is usually the case with Python libraries, \texttt{pyunicorn} is designed to 
provide simple interfaces and clear architecture, while delegating the heavy 
lifting to specialized tools for performance. Basic network measures and 
generative models are inherited from \texttt{igraph}. Wherever possible, 
numerically intensive computations are expressed as combinations of highly 
optimized linear algebra methods from \texttt{Numpy} and \texttt{Scipy}, and 
otherwise implemented in embedded \texttt{Cython}~\citep{Behnel2011} code. Thus 
all costly computations are performed in compiled C, C++ or FORTRAN code. 
Parallelization is mostly not implemented on the algorithm level, but can be 
achieved using the built-in \texttt{MPI} helper for repetitive tasks, e.g. 
computing measures on recurrence networks for different time windows of an 
observable. As the required RAM size is mostly dependent on the volume of data to be 
analyzed, a modern laptop processor with a single core suffices to perform most 
of the computations described later on for currently typical data sets in a 
matter of seconds to an hour. As an example, the recurrence network displayed in Fig.~\ref{fig:rn_lorenz} 
takes approximately 0.03 seconds to compute on a dual-core Intel Core i5 CPU with 2.4 
GHz running Mac OS X. For illustration, a more systematic study of the performance of recurrence network construction,
a common task of using \texttt{pyunicorn}, is displayed in Fig.~\ref{fig:rn_stats}.
GPU computations are currently not supported.

\section{Complex network analysis}
\label{sec:networks}

\texttt{pyunicorn} provides methods for analyzing and modeling various types of complex networks, including general networks (Sect.~\ref{sec:networks_basics}), spatial networks (Sect.~\ref{sec:spatial_networks}), networks of interacting networks (Sect.~\ref{sec:interacting_networks}) and node-weighted networks (Sect.~\ref{sec:networks_nsi}). In the following, the corresponding classes and methods are described together with a brief introduction of the underlying theory. Selected use cases illustrate the associated functionality of \texttt{pyunicorn}.

\subsection{General complex networks}
\label{sec:networks_basics}

The class \texttt{Network} in the submodule \texttt{core} serves as a parent to all other network-related classes in \texttt{pyunicorn} (see, e.g. Fig.~\ref{fig:uml}) and represents general undirected and directed \emph{networks} or \emph{graphs} $G=(V,E)$ consisting of a set of \emph{nodes} $V=\{1,\dots,N\}$ and a set of (directed) \emph{links} $E \subseteq V \times V$ without dublicates. Networks of this type can be described by an \emph{adjacency matrix} $\mathbf{A}$ with elements 
\begin{align}
A_{pq} = \begin{cases} \label{eq:adjacency}
    1 & (p,q) \in E\\
    0 & \text{otherwise}.
\end{cases}
\end{align}
Hence, $A_{pq}=1$ iff nodes $p$ and $q$ are connected by a (directed) link and $A_{pq}=0$ iff they are unconnected. In \texttt{pyunicorn}, instances of the \texttt{Network} class can be initialized using such dense adjacency matrices, but also based on sparse matrices or link lists. Link and node weights (see Sect.~\ref{sec:networks_nsi}) can be represented by link and node attributes and are accessible through the \texttt{Network.set\_link\_attribute} and \texttt{Network.set\_node\_attribute} methods, respectively. 

Many standard complex network measures, network models and algorithms are supported, e.g. degree, closeness and shortest-path betweenness centralities, clustering coefficients and transitivity, community detection algorithms and network models such as Erd\H{o}s-R\'enyi, Barabasi-Albert or configuration model random networks~\citep{Cohenbook2010,Newmanbook2010}. Several of these measures provided by \texttt{pyunicorn} can take into account directed links and link weights (directed and weighted networks) if this information is present. However, the remainder of this article focusses on undirected networks, reflecting the current state of the \texttt{pyunicorn} implementation.

Moreover, a number of less common network statistics such as Newman's~\citep{Newman2005} or Arenas'~\citep{Arenas2003} random walk betweenness can be computed. Reading and saving network data from and to many common data formats such as GraphML~\citep{Brandes2002} is possible for storage and information exchange with other software for network analysis and visualization~\citep{Nocke2015} such as \texttt{networkx}~\citep{Schult2008exploring} or \texttt{gephi}~\citep{ICWSM09154}.

\subsection{Spatially embedded networks}
\label{sec:spatial_networks}

\begin{figure}[t]
\centering
\includegraphics[width=.9\linewidth]{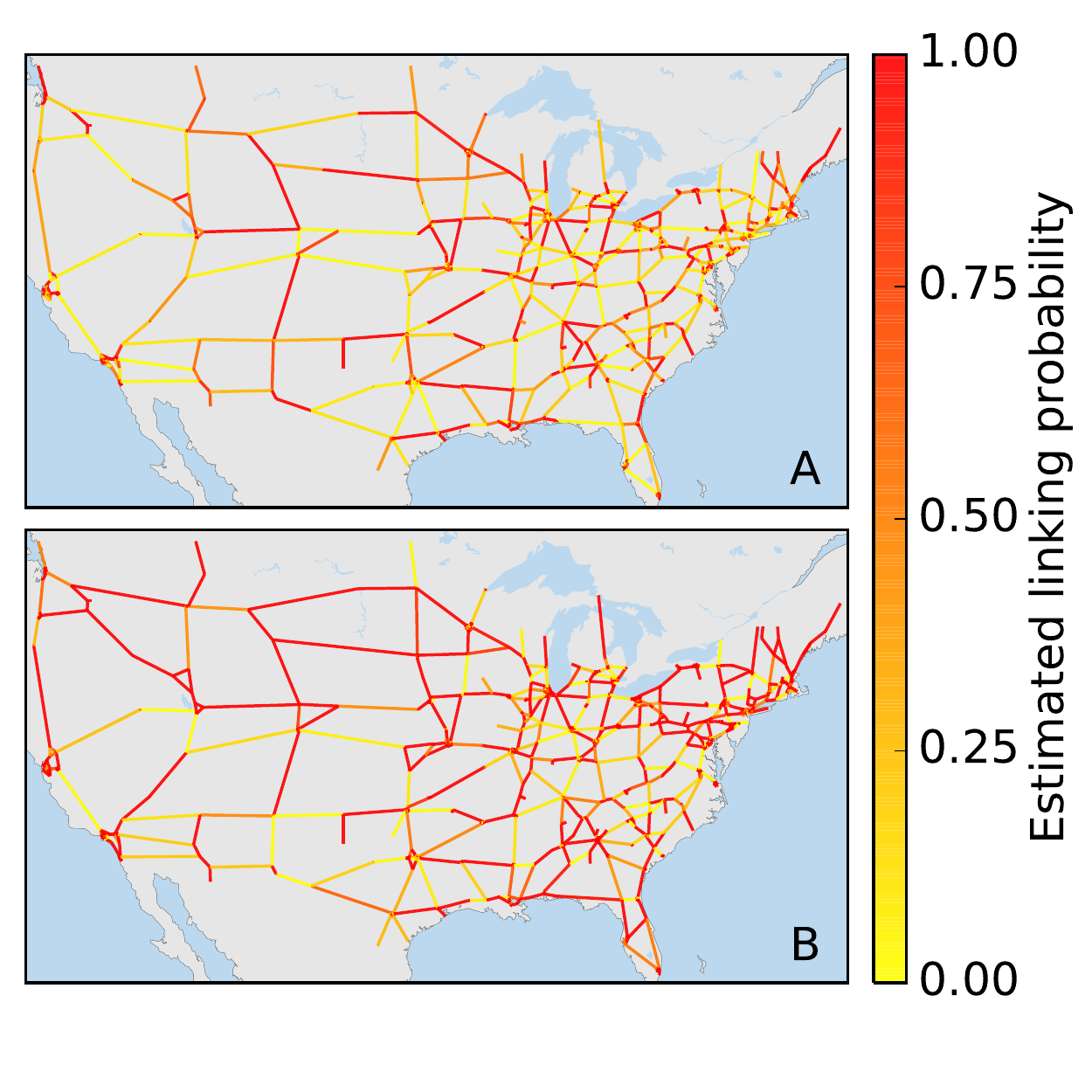} 
\caption{Effects of spatial embedding of nodes on network structure: (A) Probability for links in the US interstate network to be also present under the application of \textit{GeoModel1} computed over an ensemble of
	$100$ surrogate networks using 
	\texttt{GeoNetwork.randomly\_rewire\_geomodel\_I}. (B) The same under the application
	of \textit{GeoModel2} provided by 
	\texttt{GeoNetwork.randomly\_rewire\_geomodel\_II}.}
\label{fig:spatial_net}
\end{figure}

Many, if not most, complex networks of interest are spatially embedded~\citep{Barthelemy2011}. Consider, for example, social networks, infrastructure networks such as the internet, road and other transportation networks (Fig.~\ref{fig:spatial_net}) or functional networks in neuroscience and climatology (Fig.~\ref{fig:climate_network_scheme}). \texttt{pyunicorn} includes measures and models specifically designed for
\emph{spatially embedded networks} (or simply spatial networks) via the \texttt{GeoNetwork} class 
that inherits from the \texttt{Network} class (Fig.~\ref{fig:uml}B).
Characteristics of the nodes' spatial embedding, such as all longitudinal and
latitudinal coordinates, are stored in the \texttt{Grid} class. In
particular, this class then provides methods for computing and evaluating
spatial distances between all pairs of nodes via the methods
\texttt{Grid.angular\_distance}, \texttt{Grid.euclidean\_distance} and
\texttt{Grid.geometric\_distance\_distribution}. Additionally, functionality for
loading and saving the grid from and to common file formats such as ASCII is provided.

\subsubsection{Measures and models for spatial networks}

Measurements on the network's spatial embedding are performed by using
the class \texttt{GeoNetwork} which is initialized with an existing instance of
the \texttt{Grid} class (Fig.~\ref{fig:uml}B). Generally, all standard network measures, like the
degree or clustering coefficient, can be computed in an area-weighted variant
taking into account the network's spatial embedding and, hence, avoiding biases 
caused by the potentially widely different surface areas or volumes that nodes may represent (see Sec.~IID for details).
In addition, the distribution of the links' spatial lengths are evaluated
using the method \texttt{GeoNetwork.link\_distance\_distribution}. 
For each node in the network the lengths of its emerging links can be assessed
via the methods \texttt{GeoNetwork.average\_link\_distance},
\texttt{GeoNetwork.total\_link\_distance} and \texttt{GeoNetwork.max\_link\_distance}
which all give a notion of the spatial distance between a specific node and its
neighbors. In the application to climate
sciences, where links in the network typically represent interdependencies of statistical
significance between climate observables taken at different locations on the
Earth's surface, the above mentioned measures are of crucial importance when investigating
the presence of long-ranging teleconnections
\citep{Donges2009b,Tsonis2004,Tsonis2008role} in the
climate network (Sect.~\ref{sec:climate_networks}).

In addition to the evaluation of a spatially embedded network's
topological structure, the \texttt{GeoNetwork} class also provides random
network models to construct spatially embedded networks under the same spatial
constraints, i.e. with the same spatial distribution of nodes, as the network
under study. These spatial network surrogates allow to assess to what degree 
certain properties of an observed network are consistent with those expected 
from a structural null model that is encoded in the construction rules for the network
surrogates. In particular, the method \texttt{GeoNetwork.set\_random\_links\_by\_distance} constructs a random
network in which the probability for the presence of a link between two nodes decays
exponentially with the geographical distance between them. Furthermore,
three different \textit{GeoModels} are implemented in \texttt{pyunicorn} which
construct random network surrogates of a given network by
iteratively rewiring its links under different conditions: (i)~\textit{GeoModel1}
(\texttt{GeoNetwork.randomly\_rewire\_geomodel\_I})
creates a random network with the same link-length distribution and link density as the one
represented by the respective instance of \texttt{GeoNetwork}. (ii)~\textit{GeoModel2} (\texttt{GeoNetwork.randomly\_rewire\_geomodel\_II}) additionally preserves the local link-length distribution and degree of each node, and (iii)~\textit{GeoModel3}
(\texttt{GeoNetwork.randomly\_rewire\_geomodel\_III}) additionally sustains the
degree-degree correlations (or assortativity) of the original network~\citep{Wiedermann2015b}.

\subsubsection{Use case: US interstate network}

We illustrate the application of these random spatial network models by constructing
$100$ surrogate networks of the US interstate network
\citep{gastner_spatial_2006} utilizing \textit{GeoModel1} and
\textit{GeoModel2}, respectively.  One way to quantify how well the
network under study is represented by each of the two models is to compute the
probability that each of its links is also present in the ensemble of random surrogates (Fig.~\ref{fig:spatial_net}). We find that \textit{GeoModel1} already reproduces
well many of the original links in the US interstate network (Fig.~\ref{fig:spatial_net}A), implying that its structure is already well determined by its global link length distribution. Additionally preserving the
local link length distributions and degree sequence improves the results further: most links
of the original network are present in the surrogate networks with high
probability (Fig.~\ref{fig:spatial_net}B).

\subsection{Networks of interacting networks}
\label{sec:interacting_networks}

\begin{figure}[tbh]
\begin{center}
\includegraphics[width=0.8 \columnwidth]{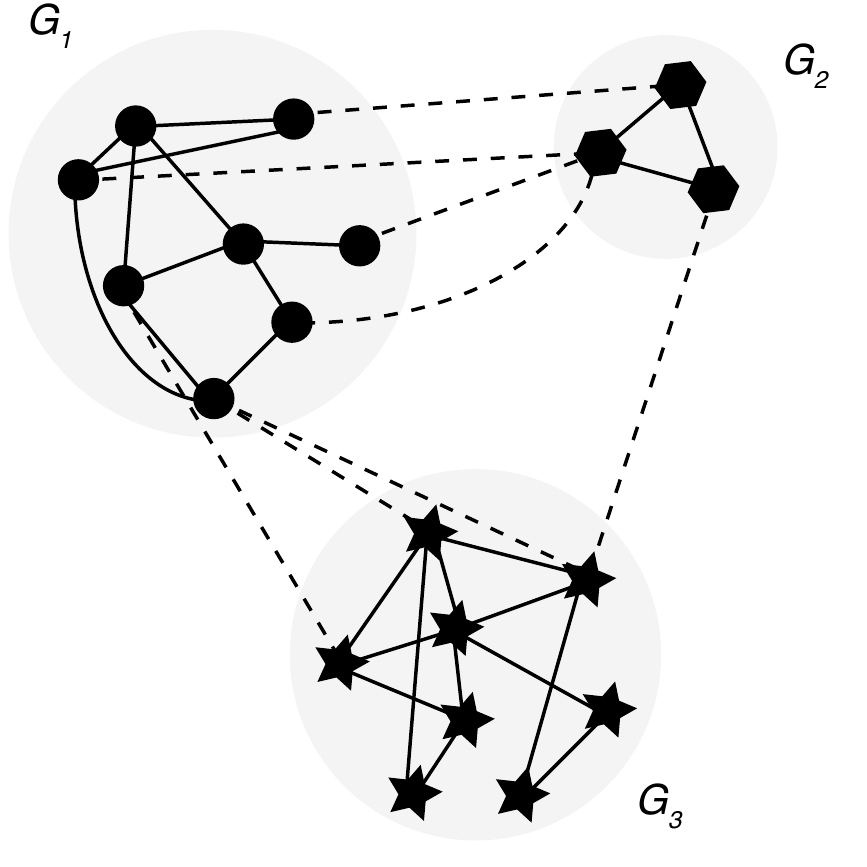}
\end{center}
\caption[]{
A network of interacting networks consisting of three subnetworks $G_i$ that can be represented and analyzed using the class \texttt{core.InteractingNetworks}. Nodes are represented by geometric symbols, while internal and cross-links are indicated by solid and dashed lines, respectively.
}
\label{fig:net_of_net}
\end{figure}

The structure of many complex systems can be described as a \emph{network of interacting or interdependent networks}~\citep{Donges2011a,Wiedermann2013}, e.g. the densely entangled infrastructures for communication and energy distribution~\citep{buldyrev2010catastrophic,gao2012networks}. Constituting a specific but important subclass of multiplex or multilayer networks~\citep{boccaletti2014structure}, these networks of networks can be represented by decomposing a network $G=(V,E)$ into a collection of $M$ \emph{subnetworks} $G_i = (V_i, E_{ii})$ (Fig.~\ref{fig:net_of_net}). Here, $V_i$ denote the disjunct sets of nodes corresponding to each subnetwork and the internal links sets $E_{ii}$ contain information on the connections within a subnetwork such that $\bigcup_{i=1}^M V_i = V$. Additionally, disjunct sets of cross-links $E_{ij}$ connect nodes in different subnetworks with $\bigcup_{i,j=1}^M E_{ij} = E$. Alternatively, a network of networks (multiplex network) of this type can be represented by a standard adjacency matrix $\mathbf{A}$ with block structure~\citep{Donges2011a}. Depending on the problem at hand, the decomposition into subnetworks can be given \textit{a priori}, as in the example of interdependent communication and electricity grids, or may be obtained from solutions of a community detection algorithm applied to a complex network of interest~\citep{Girvan2002,Newman2006,fortunato2010community}.

\subsubsection{Measures and models for networks of networks}

The \texttt{InteractingNetworks} class in the submodule \texttt{core} provides a rich collection of network measures and models specifically designed for investigating the structure of networks of networks~\citep{Donges2011a,Wiedermann2013}. Relevant examples include the \emph{cross-link density} of connections between different subnetworks (\texttt{InteractingNetworks.cross\_link\_density})
\begin{align}
\rho_{ij} = \frac{|E_{ij}|}{|V_i||V_j|},
\end{align}
the \emph{cross-degree} or number of neighbors of node $v \in V_i$ in a different subnetwork $G_j$ (\texttt{InteractingNetworks.cross\_degree})
\begin{align}
k_v^{ij} = \sum_{q \in V_j} A_{vq},
\end{align}
or the \emph{cross-shortest path betweenness} (\texttt{InteractingNetworks.cross\_betweenness}) defined for all nodes $w \in V$:
\begin{align}
b_w^{ij} = \sum_{p \in V_i, q \in V_j; p,q \neq w} \frac{\sigma_{pq}(w)}{\sigma_{pq}}
\end{align}
quantifying the importance of nodes for mediating interactions between different subnetworks, where $\sigma_{pq}$ denotes the total number of shortest paths from $p \in V_i$ to $q \in V_j$ and $\sigma_{pq}(w)$ counts the number of shortest paths between $p$ and $q$ that include $w$. The \texttt{InteractingNetworks} class also contains node-weighted versions of most of the provided statistical measures (see Sect.~\ref{sec:networks_nsi}). 

Surrogate models of interacting networks allow the researcher to assess the degree of organization in the cross-connectivity between subnetworks and its effects on other network properties of interest such as (cross-) clustering and (cross-) transitivity or shortest-path based measures such as (cross-) average path length or (cross-) betweenness~\citep{Donges2011a}. Specifically, \texttt{pyunicorn} currently supports two types of interacting network models that conserve (i)~the number of cross-links or the cross-link density between a pair of subnetworks (\texttt{InteractingNetworks.RandomlySetCrossLinks}), analogously to the Erd\H{o}s-R\'enyi model~\citep{Erdos1959} for general complex networks, and (ii)~the cross-degree sequences between a pair of subnetworks (\texttt{InteractingNetworks.RandomlyRewireCrossLinks}), corresponding to the configuration model for standard networks~\citep{Newman2003}.

\begin{figure}[tbh]
\begin{center}
\includegraphics[width=\columnwidth]{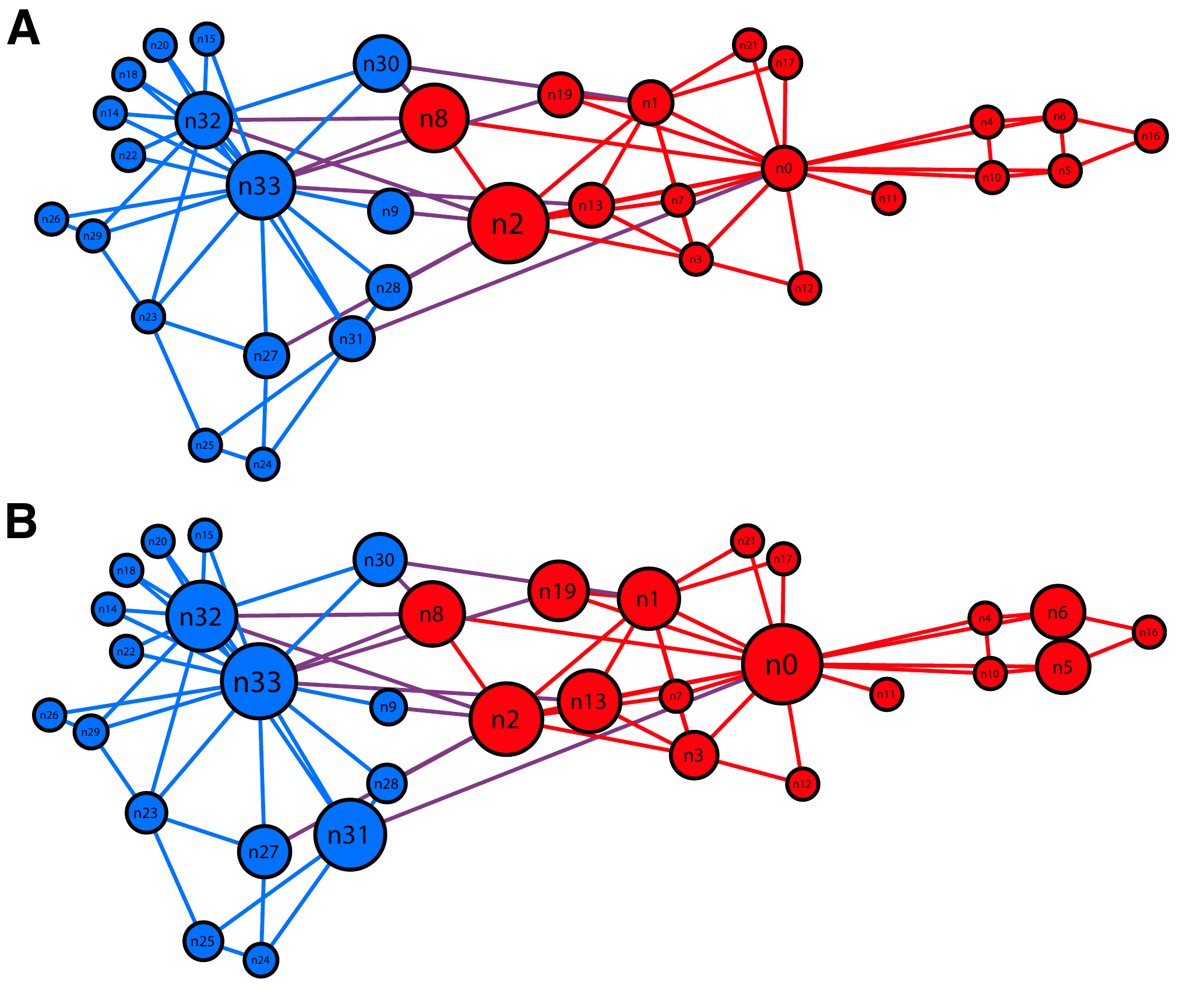}
\end{center}
\caption[]{
Visualization of the Zachary karate club network with node colors indicating the two groups emerging after fission that were lead by individuals 0 (subnetwork $G_1$, red nodes) and 33 (subnetwork $G_2$, blue nodes), respectively (note that the original node numbering of \citet{Zachary1977information} can be obtained by adding 1 to these indices). Node size is scaled according to the (A) cross-degree and (B) cross-betweenness (log-scale) measures of interacting networks analysis. Internal links are colored in red and blue, while cross-links are displayed in violet.
}
\label{fig:zachary}
\end{figure}

In the context of time series analysis, the interacting network representation has been applied for studying the structure of statistical interrelationships between different climatological fields with coupled climate networks~\citep{Donges2011a,Wiedermann2015} (Sect.~\ref{sec:coupled_climate_nets}) as well as for detecting the direction of coupling between complex dynamical systems using inter-system recurrence networks~\citep{Feldhoff2012} (Sect.~\ref{sec:recurrence}). 

\subsubsection{Use case: Zachary karate club network}

For illustrating interacting networks analysis based on a simple and commonly studied example, we choose the classical Zachary karate club social network that describes friendship relationships between 34 members of a karate club at a US university~\citep{Zachary1977information}. During the course of the study, a disagreement developed between some of the members and the club split up into two parts. Here, we represent the groups after fission by two subnetworks $G_1$ (lead by individual 0) and $G_2$ (lead by individual 33) with internal and cross-links set according to the friendship ties revealed in the original study (Fig.~\ref{fig:zachary}). The groups emerging after fission are clearly reflected in the social network structure as was also found using various community detection algorithms~\citep{Girvan2002,Newman2006}. The cross-link density $\rho_{12} \approx 0.04$ is significantly smaller than the internal link densities $\rho_{11} \approx 0.26$ and $\rho_{22} \approx 0.24$, underlining the conceptual similarities between interacting network characteristics such as cross-link densities or cross-degree sequences and measures of modularity used for community detection~\citep{Newman2006}. 

Furthermore, studying local interacting network measures yields insights into the roles of nodes with respect to interactions between the two groups. For example, nodes on the interface between both groups such as individuals 2, 8 and 30 tend to have large cross-degree and cross-betweenness values compared to other nodes on the groups' peripheries, because they serve as important connectors between the groups (Fig.~\ref{fig:zachary}). Focussing on the two group leaders, it is interesting to note that individual 0 (the instructor) has a low cross-degree (just one cross-link to $G_2$, $k_0^{1\,2}=1$) compared to individual 33 (the administrator, $k_{33}^{2\,1}=3$), while both leaders have comparable first and second largest values of cross-betweenness, $b_0^{1\,2} \approx 148$ and $b_{33}^{1\,2} \approx 94$, respectively. This observation indicates that cross-betweenness is a more reliable indicator for leadership with respect to the groups' interaction structure in this case.

\subsection{Node-weighted networks and node-splitting invariance}
\label{sec:networks_nsi}

The nodes of many real-world networks, e.g. firms, countries, grid cells, brain regions etc.,
are of heterogeneous size, represent different shares of an underlying complex system,
or are of different prior relevance to the research question at hand. As a specific example, 
in climate networks on regular latitude-longitude grids (see below and Fig.~\ref{fig:nsi}),
nodes in polar regions represent a significantly smaller fraction of the Earth's surface than do nodes in the tropics. 
If this heterogeneity can be expressed in a vector of \textit{node weights} $w_v\ge 0$ with $v \in V$,
a node-weighted network analysis seems appropriate.
Because many complex systems allow for network representations of different granularity,
the results of such a node-weighted network analysis should be consistent across scales~\citep{Heitzig2011}. \texttt{pyunicorn}  provides
node-weighted variants of most standard and many non-standard measures for networks (\texttt{Network} class)
as well as interacting networks (\texttt{InteractingNetworks} class).

\subsubsection{Measures for node-weighted networks}

The theory of \emph{node-splitting invariant} (n.s.i.) network measures 
\citep{Heitzig2011,Rheinwalt2012,Radebach2013,Wiedermann2013,Molkenthin2014,Zemp2014a,Zemp2014b,Feldhoff2015,Lange2015,Rheinwalt2015} has derived variants of many classical network measures that take into account node weights in a consistent way.
For example, the \textit{n.s.i.\ adjacency matrix $A^+_{pq}$, degree $k^\ast_v$, local and global clustering coefficients $\mathcal{C}^\ast_v, \mathcal{C}^\ast$} are defined as
\begin{align}
    A^+_{pq} &= A_{pq} + \delta_{pq}, &
    \mathcal{C}^\ast_v &= \frac{\sum_{p,q \in V} A^+_{vp} w_p A^+_{pq} w_q A^+_{qv}}{(k^\ast_v)^2},\nonumber \\
    k^\ast_v &= \sum_{p \in V} A^+_{vp} w_p, &
    \mathcal{C}^\ast &= \sum_{v \in V} w_v \mathcal{C}^\ast_v / \sum_{v \in V} w_v.
\end{align}
In contrast to their unweighted counterparts, 
these and all other n.s.i.\ measures have the following consistency property:
When a node $v$ and its weight $w_v$ are split into two interlinked nodes $v',v''$ with weights $w_{v'}+w_{v''}=w_v$ 
that are connected to the same nodes as $v$ was,
then all n.s.i.\ measures of nodes other than $v$, $v'$, and $v''$ 
remain unchanged
(e.g. the n.s.i.\ clustering coefficient of some neighbour $p$ of $v$ remains unchanged
while the ordinary clustering coefficient of $p$ would increase).
This scale consistency comes at the price that in the special case
where all weights are equal to unity, $w_v\equiv 1$, 
n.s.i.\ measures do not simply reduce to their unweighted counterparts 
but return slightly different values. 
For this reason, there exist also corrected n.s.i.\ measures 
that additionally take into account an overall typical weight $\omega\ge 0$
and have the property that in the special case where all node weights equal $\omega$,
the corrected n.s.i.\ measure equals its unweighted counterpart~\citep{Heitzig2011} .
For example, the \textit{corrected n.s.i.\ degree} $k^{\ast\omega}_v$ and \textit{local clustering coefficient} $\mathcal{C}^{\ast\omega}_v$ are
\begin{align}
    k^{\ast\omega}_v &= \textstyle \frac{k^\ast_v}{\omega} - 1, &
    \mathcal{C}^{\ast\omega}_v &= \textstyle
            \frac{\sum_{p,q \in V} A^+_{vp} w_p A^+_{pq} w_q A^+_{qv}
            -3 k^{\ast\omega}_v-1}{k^{\ast\omega}_v (k^{\ast\omega}_v-1)}.
\end{align}

In \texttt{pyunicorn}, n.s.i.\ measures are available in the \texttt{Network} and \texttt{InteractingNetworks} classes (method prefix \texttt{nsi\_}), e.g.:
\begin{verbatim}
    nsi_arenas_betweenness,
    nsi_average_neighbors_degree,
    nsi_average_path_length,
    nsi_betweenness,
    nsi_closeness,
    nsi_degree.
\end{verbatim}
Their syntax is the same as that of their unweighted counterparts, 
and some have an additional optional keyword parameter \texttt{typical\_weight}
via which the corrected n.s.i.\ measures can be requested.
To use these methods, one has to provide node weights before, 
either manually via the \texttt{Network} property
\texttt{node\_weights},
the keyword parameter \texttt{node\_weights} of the class constructor,
or automatically via the keyword parameter \texttt{node\_weight\_type} 
of the constructor of the derived class \texttt{GeoNetwork}.

\subsubsection{Use case: spatial network structures in polar region}

\begin{figure}[tbp]
  \begin{center}
      \includegraphics[width=\columnwidth]{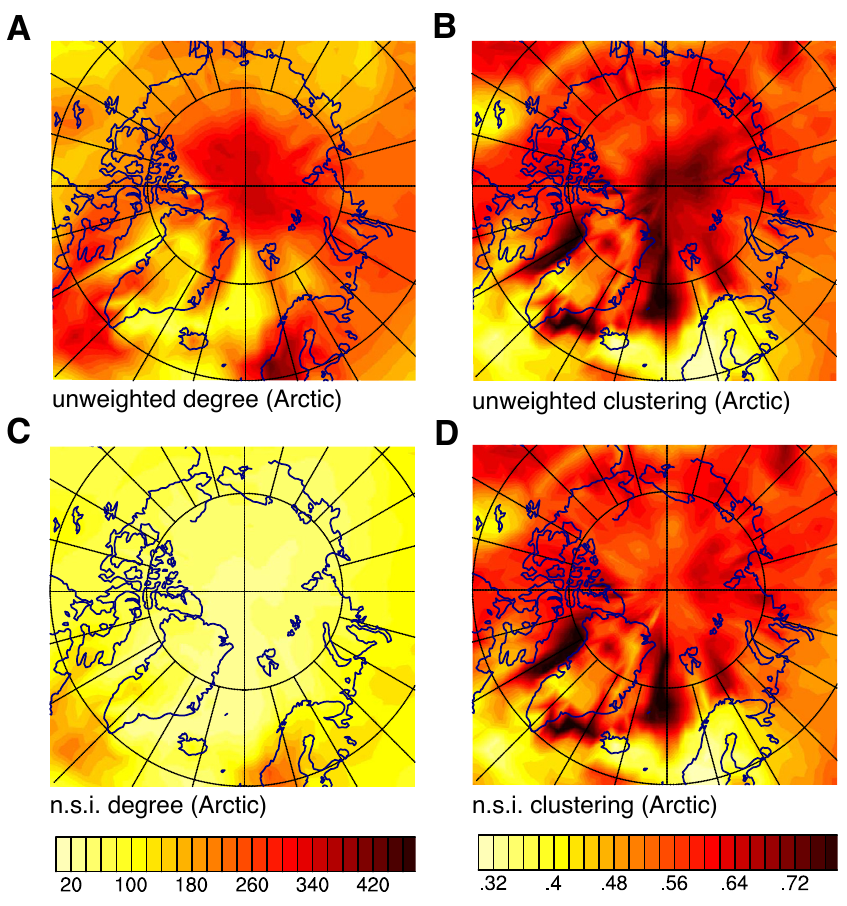}
  \end{center}
  \caption[]{\label{fig:nsi}
    Comparison of unweighted and weighted (n.s.i.) versions of
    degree (A,C) and local clustering coefficient (B,D)
    in the northern polar region (Lambert equal area projection)
    of a global climate network representing correlations in temperature dynamics.
    The high values at the pole in (A,B) turn out to be an artifact of the increasing grid density toward the pole,
    as demonstrated by (C,D).
    Reproduced with permission from Eur. Phys. J. B 85, 38 (2012) \citet{Heitzig2011}. Copyright 2012 European Physical Society and Springer-Verlag. 
  }
\end{figure}

Figure~\ref{fig:nsi} presents an application of n.s.i.\ degree and local clustering coefficient
to the functional climate network of surface air temperature dynamics in the northern polar region~\citep{Heitzig2011}.
A \texttt{ClimateNetwork} object \texttt{net} was generated as described in Sect.~\ref{sec:climate_networks},
with nodes placed on a regular latitude-longitude grid on the Earth's surface.
To reflect that grid's varying node density depending on latitude, 
the node weights were set to the cosine of latitude
by using the \texttt{ClimateNetwork} constructor's keyword parameter \texttt{node\_weight\_type="surface"} 
(inherited from \texttt{GeoNetwork}, see Fig.~\ref{fig:uml}B).
Then all nodes' n.s.i.\ degree and local clustering coefficient were computed via
\begin{verbatim}
    kstarvector = net.nsi_degree()
    Cstarvector = net.nsi_local_clustering()
\end{verbatim}
and plotted using the package \texttt{matplotlib}.
When comparing the resulting node-weighted measures (Fig.~\ref{fig:nsi}C,D) to the unweighted degree and local clustering coefficient (Fig.~\ref{fig:nsi}A,B), one realizes that the latter measures' high values around the pole (dark spots) are actually artifacts 
of the relatively higher node density, an effect that is compensated for in the n.s.i.\ measures.

\section{Functional networks: construction and analysis}
\label{sec:functional_networks}

Functional networks provide a powerful generalization of standard methods of bi- and multivariate time series analysis by allowing to study the dynamical relationships between subsystems of a high-dimensional complex system based on spatio-temporal data and using the tools of network theory. \texttt{pyunicorn} provides classes for the construction and analysis of functional networks representing the statistical interdependency structure within and between sets (fields) of time series using various similarity measures such as lagged Pearson correlation and mutual information (Sect.~\ref{sec:coupling_analysis}). Building on these similarity measures, climate networks allow for the analysis of single fields of climatological time series, e.g. surface air temperature observed on a grid covering the Earth's surface (Sect.~\ref{sec:climate_networks}). Moreover, coupled climate networks focus on studying the interrelationships between two or more fields of climatological time series, e.g. sets of time series capturing sea surface temperature and atmospheric geopotential height variability (Sect.~\ref{sec:coupled_climate_nets}). \emph{Functional network analysis} is illustrated drawing upon several examples from climatology, including the detection of spatio-temporal regime shifts in the atmosphere and ocean. While \texttt{pyunicorn} provides some functionality specific to climate data (such as the \texttt{climate.ClimateData} class), the methods for general functional network analysis can also be applied to other sources of time series such as general fluid dynamical and pattern formation systems~\citep{Molkenthin2014a}, neuroscience (e.g. functional magnetic resonance imaging, fMRI, and electroencephalogram [EEG] data; \citet{Bullmore2009}) or finance (e.g. stock market indices; \citet{huang2009network}).

\subsection{Coupling analysis}
\label{sec:coupling_analysis}

The \texttt{timeseries.CouplingAnalysis} class provides methods to estimate matrices of (optionally lagged) \emph{statistical similarities} $\mathbf{S}$ between time series including on the linear Pearson product-moment correlation and measures from information theory such as mutual information and extensions thereof. These matrices can be thresholded to obtain directed or undirected adjacency matrices for further network analysis with \texttt{pyunicorn}, e.g. as input to the \texttt{climate.ClimateNetwork} (Sect.~\ref{sec:climate_networks}) and \texttt{climate.CoupledClimateNetwork} (Sect.~\ref{sec:coupled_climate_nets}) classes. The similarity values can also be used in link-weighted network measures such as those provided by the \texttt{Network} class.

\subsubsection{Similarity measures for time series}

While standard measures such as the classical linear Pearson product-moment correlation are only briefly discussed, this section focusses on more innovative measures based on information theory that are provided by the \texttt{CouplingAnalysis} class. The latter include bivariate mutual information as well as its multivariate extensions allowing to reduce the effects of common drivers or indirect couplings (Fig.~\ref{fig:example}A) on estimates of information transfer between two processes $X,\,Y$ ($X$ and $Y$ represent time series at nodes $p$ and $q$, respectively).

All methods share the parameters \texttt{tau\_max}, the maximum time lag, and \texttt{lag\_mode}, which can be set to \texttt{'all'} to obtain a 3-dimensional \texttt{Numpy} array of shape $(\texttt{N},\,\texttt{N},\,\texttt{tau\_max + 1})$ containing lagged similarities between all pairs of nodes, or to \texttt{'max'} to return two $(\texttt{N},\,\texttt{N})$ \texttt{Numpy} arrays indicating the lag positions and values of the absolute similarity maxima.

\paragraph{Lagged cross-correlation}
The \emph{lagged Pearson product-moment correlation coefficient} (CC) of two zero-mean time series variables $X,\,Y$, implemented in \texttt{CouplingAnalysis.cross\_correlation}, is given by
\begin{align}
\rho_{XY}(\tau) = \frac{\left<X_{t-\tau}, Y_t\right>}{\sigma_X \sigma_Y}, \label{eq:pearson}
\end{align}
which depends on the covariance $\left<X_{t-\tau}, Y_t\right>$ and standard deviations $\sigma_X,\, \sigma_Y$. 
Lags $\tau>0$ correspond to the linear association of past values of $X$ with $Y$, and vice versa for $\tau<0$. In analogy, the auto-correlation is defined as $\rho_{YY}(\tau)$ for $\tau>0$. The choice \texttt{lag\_mode='max'} returns the value and lag at the absolute maximum for each ordered pair $(i,j)$, which can be positive or negative. CC is computed using the standard sample covariance estimator. It can be estimated for comparably small sample sizes. However, by definition, it does not allow to quantify nonlinear associations between time series and can produce misleading results in the presence of strongly nonlinear relationships.

\paragraph{Lagged mutual information}
Information theory \citep{Cover2006} provides a genuine framework to capture also nonlinear associations. While Shannon entropy \citep{Shannon1948} is a measure of the uncertainty about outcomes of one process, \emph{mutual information} (MI) is a measure of its reduction if another process is known. The Shannon-type MI can be expressed as
\begin{align}
I(X;Y) &= H(Y) - H(Y|X) = H(X) - H(X|Y), \label{eq:mi}
\end{align}
i.e. as the difference between the uncertainty in $Y$ and the remaining uncertainty if $X$ is already known (and vice versa). MI is symmetric in its arguments, non-negative and zero iff $X$ and $Y$ are statistically independent.
The lagged cross-MI for two time series, implemented in \texttt{CouplingAnalysis.mutual\_information}, is given by 
\begin{align}
I^{\rm MI}_{XY}(\tau) = I(X_{t-\tau};Y_t).
\end{align}
For $\tau>0$, one measures the information in the past of $X$ that is contained in $Y$, and vice versa for $\tau<0$. Correspondingly, the auto-MI is defined as $I(Y_{t-\tau};Y_t)$ for $\tau>0$.

Three different estimators are provided reflecting different trade-offs between number of samples required, bias and variance of the estimator, and computational requirements: 
\begin{itemize}
\item \texttt{estimator='binning'}: A very simple method is to quantize or partition the observation space into a set of bins (parameter \texttt{bins}). Here, we use equi-quantile bins where the bin intervals are chosen such that the marginal distributions are uniform \citep{Palus1996}. While this estimator is consistent for infinite sample size, for common sample sizes of the order $10^3$, many bins are not populated sufficiently resulting in heavily biased values of MI \citep{Vejmelka}. For example, for independent time series the estimated MI values do not center around zero.

\item \texttt{estimator='knn'}: A more advanced estimator for continuously valued variables, recommended here, is based on nearest-neighbor statistics \citep{Kraskov2004a}. This estimator is discussed in its conditional form below.

\item \texttt{estimator='gauss'}: If only the linear part of an association is desired, assuming a bivariate Gaussian distribution, the MI is simply given by
\begin{align} \label{eq:mi_corr}
I^{\rm Gauss}_{XY}(\tau) = -\tfrac{1}{2} \ln \left( 1-\rho_{XY}(\tau)^2 \right),
\end{align}
where $\rho_{XY}(\tau)$ is again the Pearson correlation coefficient.
\end{itemize}

\paragraph{Lagged information transfer}
While lagged MI can be used to quantify whether information in $Y$ has already been present in the past of $X$, this information could also stem from the common past of both processes and, therefore, is not necessarily a sign of a transfer of unique information from $X$ to $Y$. 
A first step toward a notion of directionality (the more demanding causality problem is discussed at the end of this section) is to assess a bivariate notion of \emph{information transfer} between two time series \citep{Runge2012b,Hlinka2013} in order to exclude this common past. Here, we consider two measures to achieve this goal, implemented in \texttt{CouplingAnalysis.information\_transfer}. These are based on \emph{conditional mutual information} (CMI) defined as
\begin{align}  \label{eq_cmi}
I(X;Y|Z) &= H(Y|Z) - H(Y|X,Z)
\end{align}
which can be phrased as the mutual information between $X$ and $Y$ that is not contained in a third, possibly multivariate variable $Z$. CMI shares the properties of MI and is zero iff $X$ and $Y$ are independent \emph{conditionally on $Z$}.

Following \citet{Runge2012b}, the \emph{bivariate information transfer to Y} (ITY), obtainable via the parameter \texttt{cond\_mode='ity'}, is defined as
\begin{align} 
\label{eq:def_ity}
I_{X \to Y}^{\rm ITY}(\tau) 
&= I(X_{t-\tau};Y_t|Y_{t-1},\ldots,Y_{t-p}).
\end{align}
It excludes the past of the `driven variable' $Y$ up to a maximum lag $p$ (parameter \texttt{past}). ITY can be seen as a lag-specific transfer entropy \citep{Schreiber2000b}. A more rigorous way to exclude commonly shared past is to additionally condition out the past of the `driver' variable $X$. The \emph{bivariate momentary information transfer} (MIT), called via \texttt{cond\_mode='mit'}, can be defined as
\begin{align} \label{eq:def_mit}
&I^{\rm MIT}_{X{\to}Y}(\tau)= \nonumber\\
&I(X_{t-\tau};Y_t|Y_{t-1},\ldots,Y_{t-p},X_{t-\tau-1},\ldots,X_{t-\tau-p})
\end{align}
The attribute \textit{momentary} \citep{Pompe2011} is used because MIT measures the information of the ``moment'' $t-\tau$ in $X$ that is transferred to $Y_t$. MIT can also be interpreted as a measure of causal strength as discussed in~\citet{Runge2012b}, where the multivariate versions of ITY and MIT are defined. On the downside, it is higher-dimensional resulting in a larger bias for the nearest-neighbor estimator.

Two estimators are available:
\begin{itemize}

\item \texttt{estimator='gauss'}:
Like in Eq.~(\ref{eq:mi_corr}), the CMI for multivariate Gaussians can be expressed in terms of the \emph{partial correlation}, where the Pearson correlation $\rho_{XY}$ is replaced by $\rho_{XY | Z}$.

\item \texttt{estimator='knn'}: A nearest-neighbor CMI estimator has been developed by \citet{Frenzel2007}. This estimator is computed by choosing a parameter $k$ (\texttt{knn}) as the number of neighbors in the joint space of $(X,\,Y,\,Z)$ around a sample at time $t$. The maximum-norm distance to the $k$-th nearest neighbor then defines a hypercube of length $2\epsilon_t$ for each joint sample. Then the numbers of points $k_{z,t}$, $k_{xz,t}$ and $k_{yz,t}$ in the subspaces with distance less than $\epsilon_t$ are counted, and the CMI is estimated as
\begin{align}
\widehat{I}(X;Y|Z) = \psi(k) + \frac{1}{T} \sum_{t=1}^T \left[ \psi(k_{z,t}) - \psi(k_{xz,t}) -\psi(k_{yz,t}) \right],
\end{align}
where $\psi$ is the Digamma function and $T$ is the number of samples. Smaller values of $k$ result in smaller cubes and, since as assumed in the estimator's derivation, the density is approximately constant inside these, the estimator has a low bias. Conversely, for large $k$ the bias is stronger, but the variance is smaller.  Note, however, that for independent processes the bias is approximately zero, i.e. $\widehat{I}(X;Y|Z)\approx 0$, and large $k$ are therefore better suited for (conditional) independence tests, e.g. on whether a link exists between two time series.

\end{itemize}

\subsubsection{Use case: coupled stochastic processes}
\begin{figure}[t!]
\begin{center}
\includegraphics[width=1.\columnwidth]{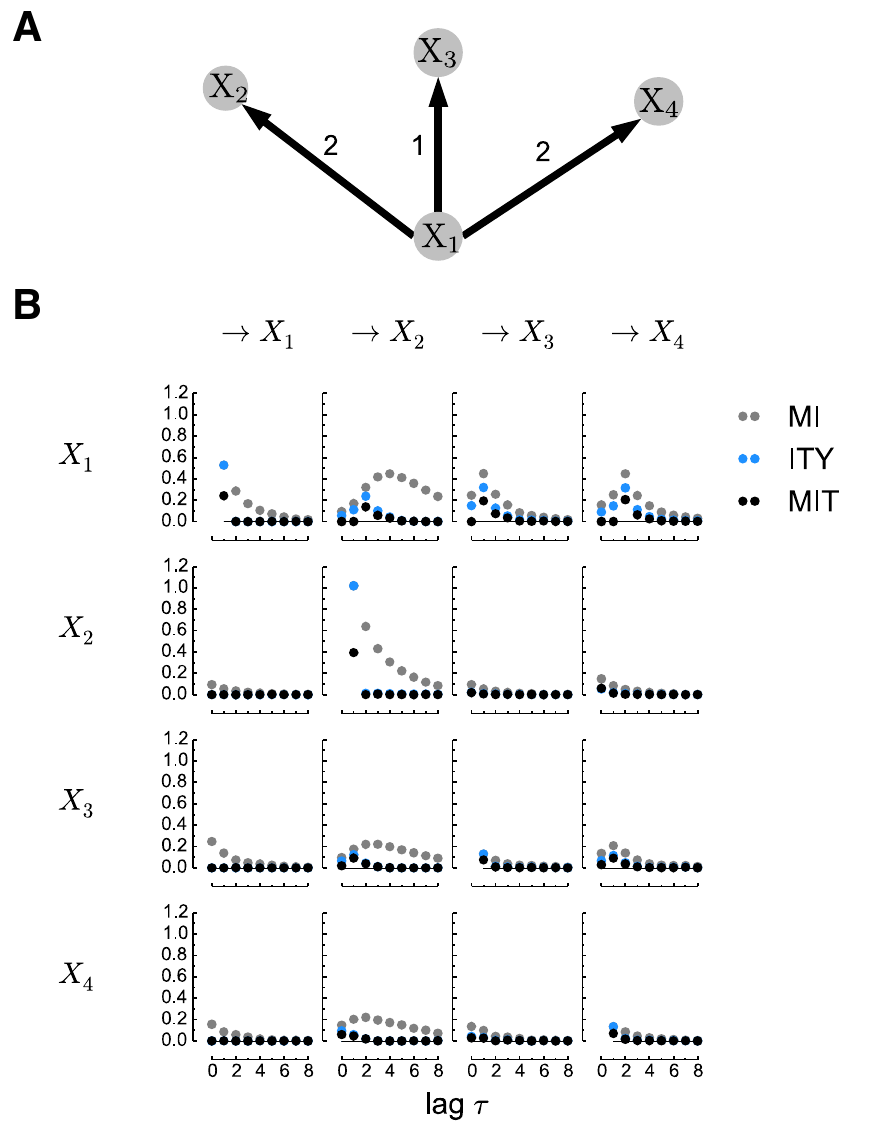}
\end{center}
\caption[]{
(A) Process graph and (B) matrix of lag functions for similarity measures MI, ITY, and MIT (the latter two with past-history parameter $p=1$) for a realization of example process~(Eq. (\ref{eq:example})). All methods implemented in \texttt{CouplingAnalysis} have a parameter \texttt{lag\_mode}, which can be set to \texttt{'all'} to return all lagged similarities between all pairs of nodes (shape $(\texttt{N},\,\texttt{N},\,\texttt{tau\_max + 1})$, or to \texttt{'max'} to return two $(\texttt{N},\,\texttt{N})$-matrices of the value at the absolute maximum of each panel and the corresponding lags. Note that this results in an asymmetric matrix which can be symmetrized by taking the maximum of each pair $(i,j)$ and $(j,i)$ with the method \texttt{CouplingAnalysis.symmetrize\_by\_absmax}.
}
\label{fig:example}
\end{figure}
Consider the following simple four dimensional process to illustrate the different measures (Fig.~\ref{fig:example}A):
\begin{align} \label{eq:example}
X_1(t) &= 0.8 X_1(t-1) + \eta_1(t) \nonumber\\
X_2(t) &= 0.8 X_2(t-1) + 0.5 X_1(t-2) +  \eta_2(t) \nonumber\\
X_3(t) &= 0.7 X_1(t-1) + \eta_3(t) \nonumber\\
X_4(t) &= 0.7 X_1(t-2) + \eta_4(t)\,,
\end{align}
where $\eta_i$ are independent zero-mean and unit variance Gaussian innovations. 
Here $X_{1,2}$ are auto-correlated and $X_1$ drives $X_2$ at a lag of 2, $X_3$ at a lag of 1, and $X_4$ at a lag of 2. Figure~\ref{fig:example}B shows the lag functions for all pairs of variables.
We illustrate in the following, how the different measures MI (similar to CC), ITY, and MIT can be used to reconstruct an adjacency matrix of a functional network. 
Regarding a directed link between $X_1$ and $X_2$, both directions $X_1\to X_2$ and $X_2\to X_1$ have non-zero MI values, making it hard to conclude on a direction. Further, the peak of the MI function in $X_1\to X_2$ is at lag $\tau=4$, even though the driving lag is only 2. The ambiguity in interpreting the value of MI is discussed in \citep{Runge2012b} and the problem that coupling delays cannot be properly inferred with MI because the peak of the lag function is shifted for strong auto-correlations is analyzed in \citet{Runge2014a}.
These shortcomings can be overcome with ITY and MIT. Both measures feature a much sharper peak at the correct lag $\tau=2$. The value of MIT is smaller because it also excludes the effect due to the auto-correlation of the driving variable.

Still, all these bivariate lagged measures show non-zero values even if NO physical coupling is present in Eq.~(\ref{eq:example}), e.g. in the lower two rows in Fig.~\ref{fig:example}B from $X_3$ and $X_4$ to the other processes. These artifacts are due to indirect links and common drivers (Fig.~\ref{fig:example}A), e.g. $X_1$ driving $X_3$, and $X_4$ leading to a spurious peak at $X_3\to X_4$. MI and the bivariate versions of ITY and MIT discussed here are also not able to reliably identify the correct coupling lags when multiple lags are present.

\subsubsection{Discussion and extensions}

Generally, networks reconstructed from bivariate similarity measures can be used to study statistical properties of ÔassociationsÕ between time series, but cannot be interpreted in a causal context. Based purely on observational data, a notion of a causal network can be defined within the framework of \emph{time series graphs} \citep{Eichler2011,Runge2012prl}, which can be efficiently estimated by causal discovery algorithms in a linear framework with partial correlation \citep{Runge2014a,Schleussner2013} or with non-parametric information-theoretic estimators as implemented in the causal algorithm proposed in \citet{Runge2012prl}. Based on these causal graphs, multivariate versions of ITY and MIT can be used to quantify the links' strength at the correct causal lag \citep{Runge2012b}. These methods are available in the software package \texttt{tigramite}, which can be obtained from the website \texttt{http://tocsy.pik-potsdam.de/tigramite.php} as a complement to \texttt{pyunicorn}. Note, however, that reliable causal analyses, especially with information-theoretic estimators, require much more samples than classical bivariate analysis, which typically restricts their applicability to much smaller networks \citep{Balasis2013}. An alternative to classical path-based network measures is discussed in \citet{Runge2014c,Runge2015} and introduces quantifiers of information transfer through causal pathways.

\subsection{Climate networks}
\label{sec:climate_networks}

\begin{figure*}[tbp]
\centering
\includegraphics[width=0.75\textwidth]{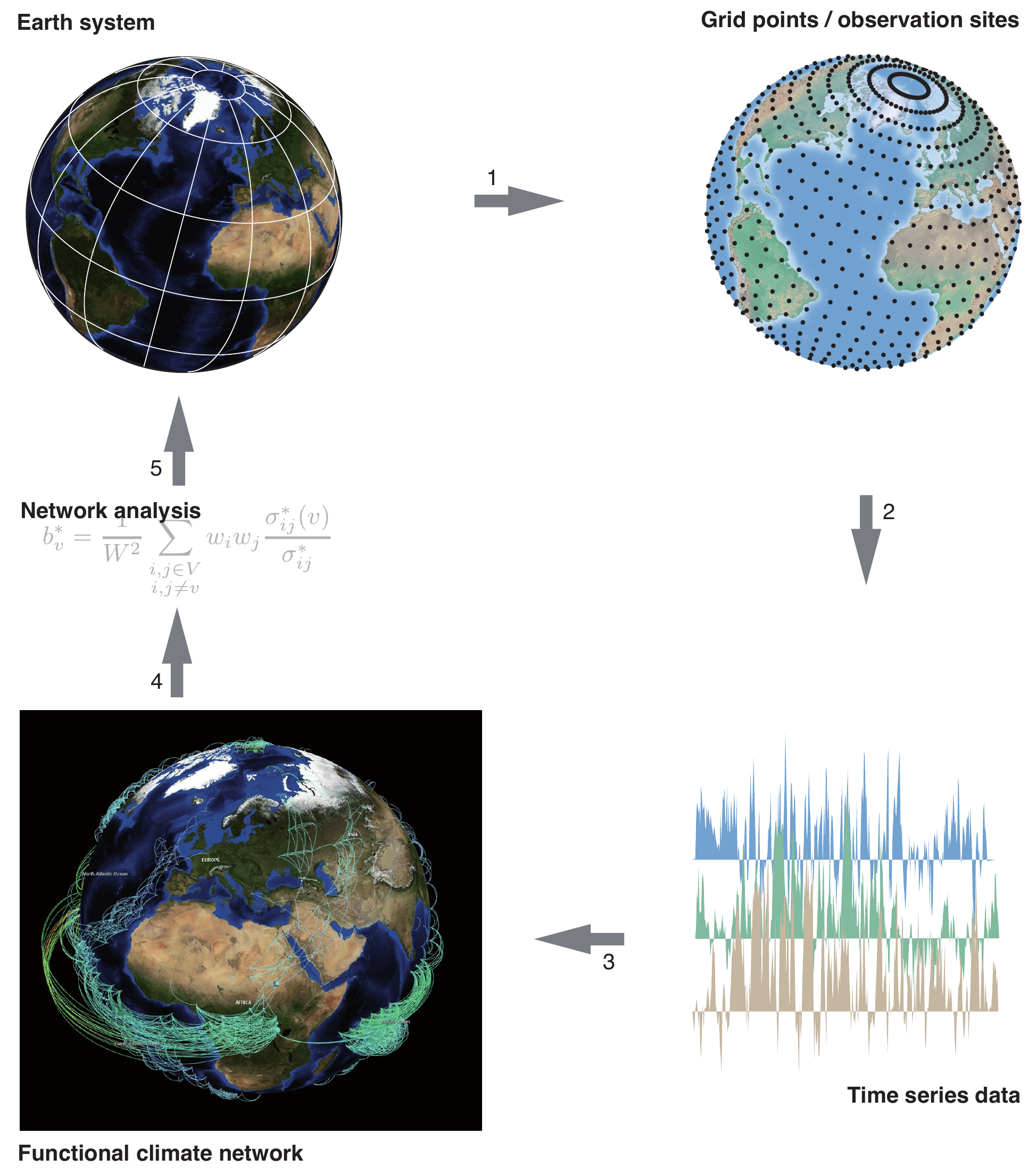}
\caption{Workflow of functional network analysis illustrated for climate networks (modified from \citet{Donges2012b}). In step 1, a discretized time series representation $\{x_v(t)\}_{v=1}^N$ of the climatological field(s) of interest is chosen that is usually prescribed by the available gridded or station data. Step 2 includes time series preprocessing and the computation of similarity measures $S_{ij}$ for quantifying statistical interdependencies between pairs of climatological time series. In step 3, the construction of a climate network from the similarity matrix $\mathbf{S}$ typically involves some thresholding criterion (see \citet{tominski2011information,Nocke2015} for details on the climate network shown here that was visualized using the software CGV~\citep{tominski2009cgv}). In step 4, the obtained climate network is investigated drawing on the tools of complex network theory. Finally, in step 5, the results of this analysis need to be interpreted in terms of the underlying dynamical Earth system.}
\label{fig:climate_network_scheme}
\end{figure*}

As a typical application of functional networks, \emph{climate network analysis} is a versatile approach for investigating climatological data and can be seen as a generalization and complementary method to classical techniques from multivariate statistics such as eigen analysis (e.g. empirical orthogonal function or maximum covariance analysis)~\citep{Donges2015}. It has been already successfully used in a wide variety of applications, ranging from the complex structure of teleconnections in the climate system~\citep{Tsonis2004,Tsonis2008role,Donges2009a}, including backbones and bottlenecks~\citep{Donges2009b,Runge2014c}, to dynamics and predictability of the El Ni\~no-Southern Oscillation (ENSO)~\citep{Radebach2013,Ludescher2013,Ludescher2014}.

Climate networks (class \texttt{climate.ClimateNetwork}) represent strong statistical interrelationships between time series and are typically reconstructed by thresholding the matrix of a statistical similarity measure $\mathbf{S}$ (Fig.~\ref{fig:climate_network_scheme}) such as those derived from coupling analysis (Sect.~\ref{sec:coupling_analysis}):
\begin{equation}
A_{pq} = \begin{cases}
    \Theta\left(S_{pq} - \beta \right) & \text{if}\,\,\, p \neq q,\\
    0 & \text{otherwise},
\end{cases}
\end{equation}
where $\Theta(\cdot)$ is the Heaviside function, $\beta$ denotes a threshold parameter, and $A_{pp}=0$ is set for all nodes $p$ to exclude self-loops. The threshold parameter can be fixed following considerations of statistical significance given a prescribed null hypotheses (\texttt{ClimateNetwork.set\_threshold}), set individually to $\beta_{pq}$ for each pair of time series or chosen to achieve a desired link density in the resulting climate network (\texttt{ClimateNetwork.set\_link\_density}). 

Certain types of time series preprocessing such as calculation of climatological anomaly values (by subtracting phase averages to reduce the first-order effects of the annual cycle) are provided by the \texttt{climate.ClimateData} class included in \texttt{ClimateNetwork}. The classes derived from \texttt{ClimateNetwork} (Fig.~\ref{fig:uml}B) apply different types of similarity measures for network construction, e.g. \texttt{TsonisClimateNetwork} for linear Pearson correlation at zero lag or \texttt{MutualInfoClimateNetwork} for nonlinear mutual information at zero lag. Note that for climate network analysis of large data sets with more than one million time series the \texttt{par$@$graph} software~\citep{Ihshaish2015} can be used, offering methods and measures comparable to that of the \texttt{TsonisClimateNetwork} class.

In the following, we present two use cases of studying fields of single climatological observables using climate networks (Sects.~\ref{sec:climate_transitions}, \ref{sec:evolving_climate_nets}) and one use case of investigating the coupling structure between two climatological fields using coupled climate networks (Sect.~\ref{sec:coupled_climate_nets}).

\subsubsection{Use case: climate networks for detecting climate transitions}
\label{sec:climate_transitions}

Climate networks (CNs) based on spatial correlations of time series 
have recently been introduced to develop early warning  indicators for 
climate transitions.  Two types of CNs have mainly been used: Pearson Correlation 
Climate Networks (PCCN), where the Pearson correlation (Eq.~\ref{eq:pearson}) is applied to measure 
connectivity in the network, and Mutual Information Climate Networks (MICN),
where mutual information (Eq.~\ref{eq:mi}) is applied \citep{Feng2014a}.  PCCNs and MICNs can 
be reconstructed by the \texttt{pyunicorn} classes \texttt{climate.TsonisClimateNetwork} 
and \texttt{climate.MutualInfoClimateNetwork}, respectively.  

\begin{figure*}[tbp]
\begin{center}
	\includegraphics[width=1\textwidth]{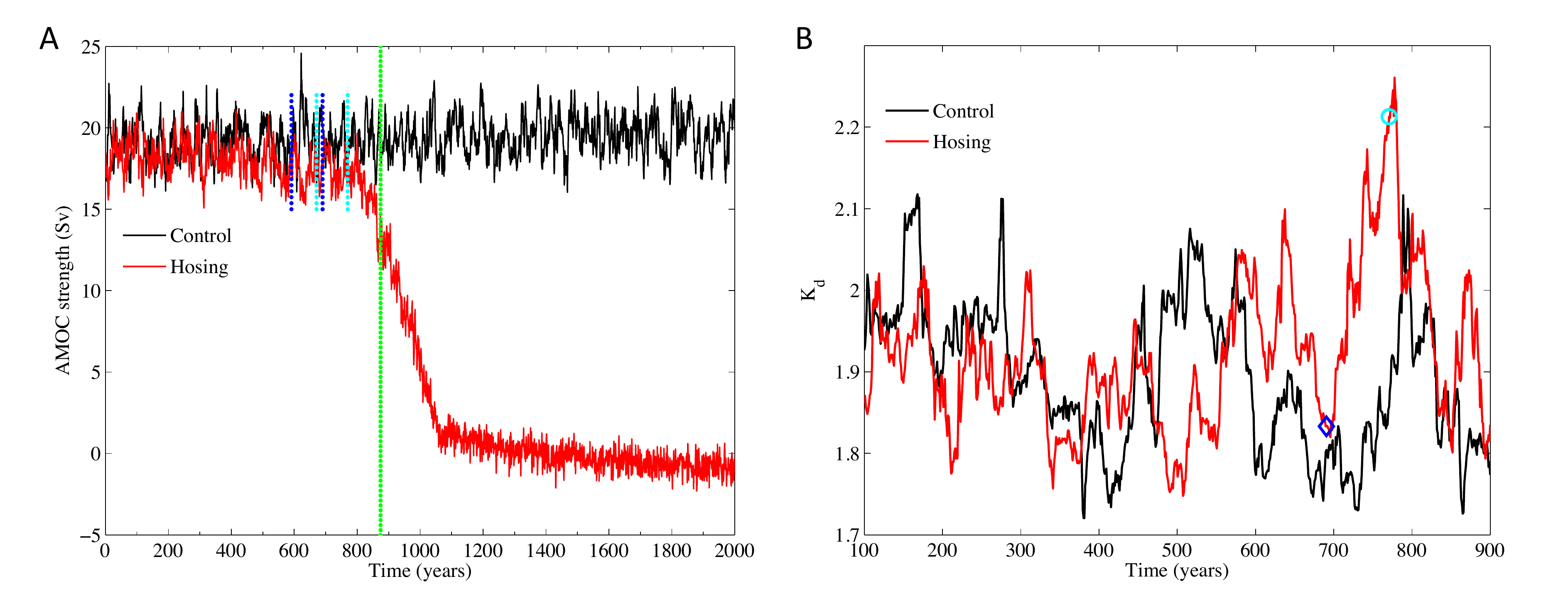} 
\end{center}
\caption{(A) Time series of the annual mean MOC (in Sv, 
1 Sv = 10$^6$ m$^3$s$^{-1}$)   at 26$^\circ$N and 1000m depth 
in the Atlantic  for the control simulation  (black curve) and the 
hosing simulation (red curve) of the FAMOUS  model. The green 
line indicates the collapse time $\tau_c = 874$ years. (B) The kurtosis 
indicator  $K_d$ gives an early warning signal of the collapse  at 738 
years and lasts  for 44 years. The blue diamond marker shows the 
$K_d$ value for the window of years 591--690 as indicated by blue dashed 
lines in panel (A), and the cyan marker shows $K_d$ for years 671--770 as indicated by cyan dashed lines in panel (A).
}
\label{f:MOCF1}
\end{figure*}

One climate transition of crucial interest is the possible collapse of  the Atlantic  Meridional Overturning 
Circulation (MOC) \citep{Mheen2013, Feng2014b} as is occurring in  
simulations of the Fast Met Office/UK Universities Simulator (FAMOUS) climate 
model \citep{Hawkins2011}. Figure~\ref{f:MOCF1}A  shows time series of annual 
mean  Atlantic MOC strength  for both the control simulation (black curve) and 
the hosing simulation  (red curve),  at the location where the maximum MOC occurs  
(at latitude 26$^\circ$N  and 1000 m depth). The hosing simulation performed using the FAMOUS 
model is a freshwater-perturbed experiment, in which the freshwater influx over 
the extratropical North Atlantic is increased  linearly  from  0.0 Sv to 1.0 Sv 
(1 Sv = 10$^6$ m$^3$s$^{-1}$) over 2000 years \citep{Hawkins2011}.  
One can see that the MOC values of the control simulation are statistically  
stationary  over the 2000-year integration  period, while the MOC values for 
the hosing  simulation  show a  rapid decrease between the years 800 and 1050. 
Based on a threshold criterium it was found that  the MOC collapses at 
$\tau_c=874$ years~\citep{Feng2014b}, as is shown by the green 
line in Fig.~\ref{f:MOCF1}A.

\begin{figure*}[tbp]
\begin{center}
	\includegraphics[width=1\textwidth]{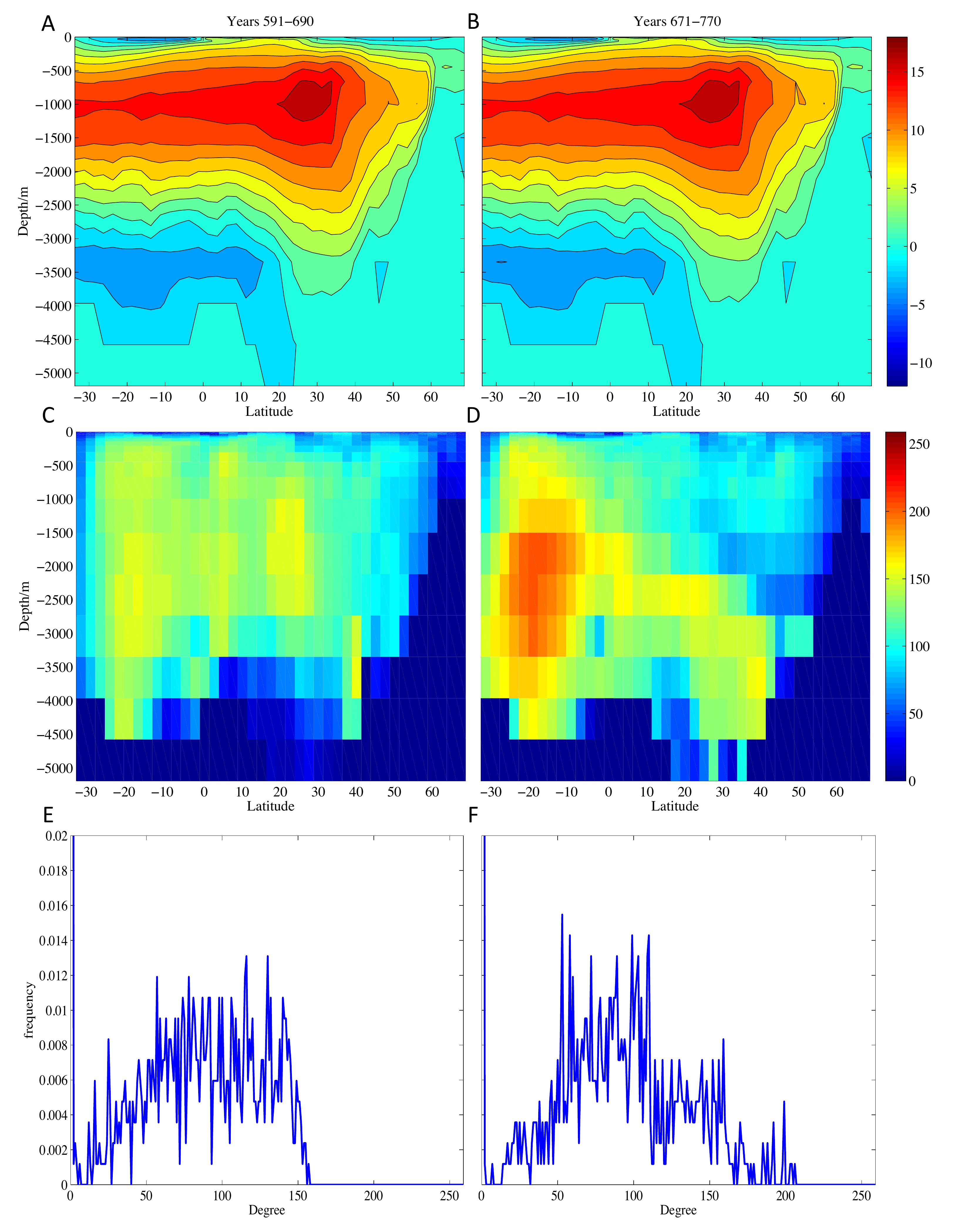} 
\end{center}
\caption{(A,B) Mean MOC stream function pattern over 
two 100 year windows (years 591--690 and years 671--770). (C,D) Degree field of the 
Pearson Correlation Climate Network (PCCN) constructed  from the MOC 
data over each window using a threshold value of  $\tau=0.5$.
(E,F) Degree frequency distribution for both cases. 
}
\label{f:MOCF2}
\end{figure*}

In \citet{Feng2014b}, PCCNs were reconstructed from the MOC field of the 
FAMOUS model using a 100-year sliding window (with a time step of one year).  It 
was found that the kurtosis $K_d$ of the climate network's degree distribution (\texttt{TsonisClimateNetwork.degree\_distribution}) is a useful
early warning indicator for the collapse. The values of  $K_d$ for the hosing 
simulation (red curve)  and for the control  simulation (black curve) are plotted  
in Fig.~\ref{f:MOCF1}B. For the hosing simulation,  there is indeed a  strong 
increase of $K_d$ significantly exceeding the values for the control 
simulation at 738 years. The classical critical slowdown indicators like variance 
and lag-1 auto-correlation  based on the same MOC data (using the same sliding 
window) do not  show any early warning signal of the MOC transition before the  
collapse  time  $\tau_c$ \citep{Feng2014b}. 

To see why the kurtosis $K_d$ of the degree distribution of PCCNs is 
an effective indicator for the Atlantic MOC collapse,  we show  in 
Fig.~\ref{f:MOCF2} the  mean MOC fields (A,B), the degree fields of the 
PCCNs (C,D) and the degree distribution (E,F) for two 100-year  windows 
(years 591--690 and years 671--770) near the transition. Although the MOC 
is gradually weakening  with the  freshwater inflow,   the changes in the MOC 
pattern are relatively minor.  However, the changes in the degree field  
are substantial and when the freshwater inflow is increased,  high degrees in 
the network appear at nodes in the deep ocean (at about 1000 m depth) at  
mid-latitudes, especially in the South Atlantic.   The histograms of the degree 
fields (the degree distributions) for  these windows (Fig.~\ref{f:MOCF2}E,F) 
show a tendency towards high degree, which is  successfully   captured by the 
kurtosis $K_d$.

The collapse of the Atlantic MOC has been identified as one 
of the important  tipping points in the climate system \citep{Lenton2008}, 
as it will lead to a significantly reduced northward heat transport \citep{Bryan1986, 
Rahmstorf2000}.  With the tools of \texttt{pyunicorn}, we have provided a novel early 
warning  indicator based on climate networks.  The particular advantage of such an indicator,
in contrast to the indicators based on a single-point time series, is that it reflects
spatial correlations.  When applied to data from the FAMOUS model, our 
results  show that  this kurtosis indicator $K_d$  provides a 
strong early warning signal at least 100 years before the transition.

\subsubsection{Use case: seasonal and evolving climate network analysis of monsoon variability}
\label{sec:evolving_climate_nets}

 
Temporal and spatial variability of climate, and thus climate network structure, are of increasing 
interest considering ongoing environmental changes. Functional climate networks evolving in time are
a promising and useful tool for analyzing spatial and temporal transitions in climate and various other
climatic phenomena \citep{Rehfeld2012,Radebach2013,Gozolchiani2008}. 
In particular, evolving climate networks have been 
used to study seasonal and annual variability of the Indian Monsoon system as one of the major global climatic subsystems affecting life and prosperity of 1/4th of the world's human population \citep{Malik2011,Stolbova2014,Tupikina2014}. 
On seasonal time scales, it is crucial to identify spatial structures of synchronicity of extreme rainfall events over 
the Indian monsoon domain, as extreme rainfall events are the main cause of devastating floods on the subcontinent.
On annual time scales, variability of the surface air temperature (SAT) field is of great interest, as it influences the total amount of rainfall and its spatial distribution during the monsoon season.

\paragraph{Data and methodology for network construction}

\begin{figure*}[tbp]
\begin{center}
\includegraphics[width=0.81 \textwidth]{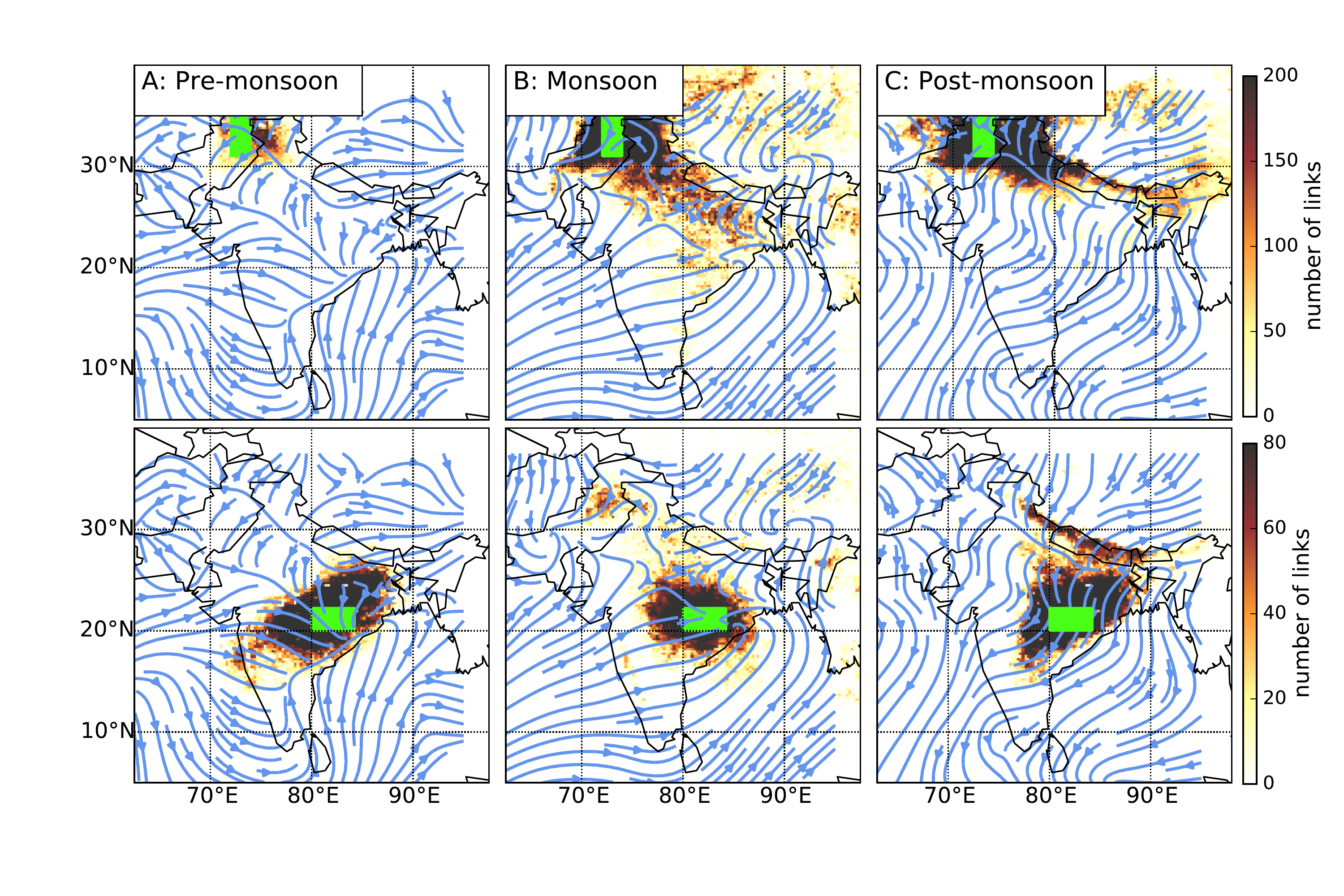}
\end{center}
\caption{Maps of the number of outgoing undirected links (indicated by color scale) from a set of 153 reference grid points (green rectangles), representing synchronization of extreme rainfall events, and the mean surface wind vector (seasonal, 1998--2012), the latter indicated by blue arrows. The reference regions are: North Pakistan (top panels) and the Eastern Ghats (bottom panels).}
\label{fig:NPG_links_TRMM}
\end{figure*}

In order to study seasonal extreme rainfall variability, we used observational satellite daily rainfall data for the period 1998 -- 2012 
(TRMM~3B42V7 \citep{Huffman2007,TRMM} with a spatial resolution of 0.25$^\circ$\,$\sim$\,25\,km, extracted for the South Asian
region (62.375--97.125$^\circ$\,E, 5.125--39.875$^\circ$\,N)). First, we defined time series of extreme rainfall events by considering daily precipitation above the 90th-percentile for each rainfall time series as extreme. Then, we constructed seasonal climate networks for three time periods: pre-monsoon (March--May), monsoon (June--September) and post-monsoon seasons (October--December) using event synchronization~\citep{Quiroga2002,Boers2014} -- a measure of synchronicity of extreme rainfall events between a pair of geographical locations (\texttt{climate.EventSynchronizationClimateNetwork} class).

For the analysis of annual SAT variability over the Asian monsoon domain we used daily temperature anomaly  
data (NCEP/NCAR reanalysis~\citep{Kistler2001} for the Asian monsoon region 57.5--122.5$^\circ$\,E, 2.5--42.5$^\circ$\,N) and construct yearly climate networks for the period 1970--2010 based on Pearson correlation at zero lag using the \texttt{climate.TsonisClimateNetwork} class. We consider a set of 40 static networks obtained from thresholded correlation matrices as one time evolving temporal climate network of the Asian Monsoon domain.

\paragraph{Temporal network measures}

For analyzing the annual variability of climate networks of the Asian monsoon domain, we use standard network measures~~\citep{Newman2003} for quantifying changes in time evolving networks~\citep{holme2012temporal}, as described in \citet{Radebach2013,Tupikina2014}. Specifically, we calculate average path length $\mathcal{L}$ (\texttt{TsonisClimateNetwork.average\_path\_length}) and transitivity $\mathcal{T}$ (\texttt{TsonisClimateNetwork.transitivity}) for each time step in the temporal climate network.

\paragraph{Results}
Analysis of seasonal networks of extreme rainfall events revealed two key regions, North Pakistan and the Eastern Ghats, which 
influence the distribution and propagation of extreme rainfall over the Indian subcontinent (Fig.~\ref{fig:NPG_links_TRMM}). The Eastern Ghats region was previously known by climatologists as an area influencing rainfall over the Indian subcontinent due to its topography, causing orographic rainfall. However, the complex climate network approach allows us to obtain new insights into the climatology of extreme rainfall events and to
detect a previously unknown influential region: North Pakistan. This finding pinpoints the strong influence of climatological
phenomena such as western disturbances on extreme rainfall events over the Indian subcontinent. It opens new possibilities to account for
North Pakistan as a key region for inferring interactions between the Indian Summer Monsoon system and western disturbances, 
and based on this information, to improve the forecasting of extreme rainfall events over the Indian subcontinent~\citep{Stolbova2014}.

\begin{figure*}[tbp]
\includegraphics[width=0.76 \textwidth]{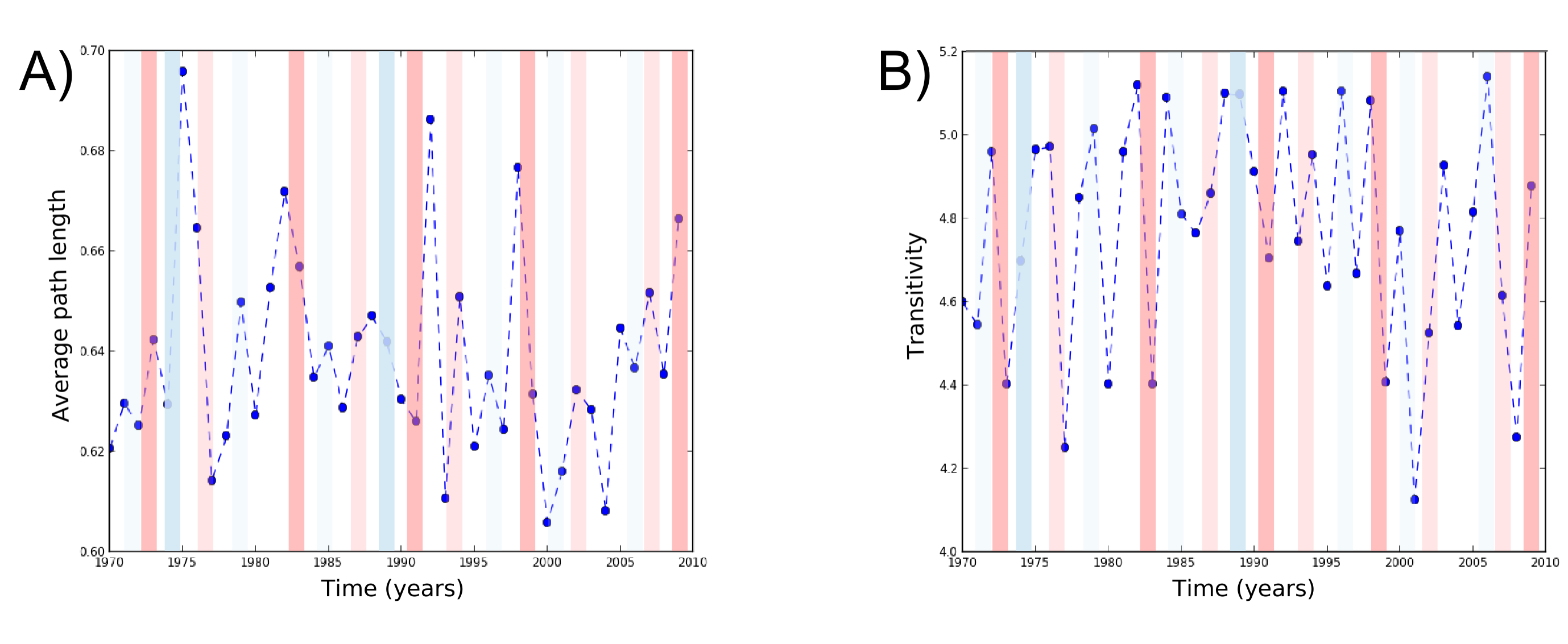} 
\caption{Evolving average path length $\mathcal{L}$ (A) and transitivity $\mathcal{T}$ (B) network measures calculated for each yearly time step in the temporal climate network constructed for the period (1970--2010). 
Most of the peaks of $\mathcal{L}$ correspond to big El Ni\~no (EN) 
years (red vertical bars), troughs of $\mathcal{T}$ correspond to La Ni\~na events (blue vertical bars) where 
color intensity of the bar corresponds to EN event strength. The data has spatial resolution $2.5^\circ\times 2.5^\circ$ covering the area between 2.5$^\circ$S to $42.5^\circ$N and 57.5$^\circ$E to 122.5$^\circ$, i.e. 468 nodes.}
\label{trans_av}
\end{figure*}

Analysis of the annual variability of the evolving Asian monsoon SAT climate network allows us to conclude that a
highly non-random, deterministic general structure is present in the network on which the inter-annual variability is imprinted \citep{Radebach2013,Molkenthin2014,Tupikina2014}. 
The annual climate network variability  
could be explained by a dominant influence of the topography of the region on the climate network as well as regular monsoon effects, 
or by dominant climatic events such as El Ni\~no or La Ni\~na \citep{Gozolchiani2008,Tsonis2008a,Berezin2012}. 
Observing the changes in temporal climate network properties such as average path length $\mathcal{L}$ and transitivity $\mathcal{T}$ allows to investigate this question further (Fig.~\ref{trans_av}). 
Most of the peaks of $\mathcal{L}$ correspond to big El Ni\~no (EN) 
years, while troughs of $\mathcal{T}$ correspond to La Ni\~na years according to classification of EN in \citet{Kug2009}. 
This coincides well with results concerning the annual variability of global temporal climate networks \citep{Radebach2013} and indicates the presence of teleconnections between El Ni\~no and Indian Monsoon region. 

\paragraph{Conclusions}

Understanding the variability and evolution of the Indian monsoon and its interactions with ENSO 
remains one of the most vital questions in climatology. Using the \texttt{pyunicorn} toolbox we were able to analyze these phenomena 
and their interactions from a climate network perspective. Following this approach revealed the influence of western disturbances and westerlies on the synchronicity, spatial structure and seasonal dynamics of extreme rainfall events over the Indian subcontinent and yielded insights into the annual evolution of temperature climate networks over the Indian monsoon domain, and the influence of ENSO on the Indian monsoon system.

\subsection{Coupled climate networks}
\label{sec:coupled_climate_nets}

Coupled climate networks~\citep{Donges2011a,Feng2012} generalize climate network analysis to the statistical interdependency structures between two or more fields of climatological observables as a network of interacting networks (Fig.~\ref{fig:net_of_net}) and, hence, provide a complementary approach and generalization of classical methods of eigen analysis such as maximum covariance analysis~\citep{Donges2015}. \texttt{pyunicorn} provides the functionality to construct and analyze coupled climate networks via the \texttt{CoupledClimateNetwork} class, which inherits from \texttt{ClimateNetwork} and \texttt{InteractingNetworks}. In accordance with the n.s.i.\ framework~(Sect.~\ref{sec:networks_nsi}), weighted versions of all measures are also available in these classes, e.g. allowing for an area-weighted computation of all interacting network
measures~\citep{Wiedermann2013}. This is particularly useful when studying coupled climate networks
that cover areas close to the poles, as in most cases the density of nodes in
these areas varies strongly due to the regular gridding of many climate
data sets (e.g. Fig.~\ref{fig:nsi}).

We have applied the coupled climate networks framework to study ocean-atmosphere
interactions in the Northern hemisphere based on the monthly HAD1SST sea-surface temperature (SST) \citep{rayner_global_2003}
and the 500 mbar geopotential height (HGT) fields from the ERA40 reanalysis project
\citep{uppala_era-40_2005} for all nodes northward of $30^\circ$N latitude and
using the linear Pearson correlation coefficient at zero lag~\citep{Wiedermann2015}. 

\begin{figure}[t]
\centering
\includegraphics[width=\columnwidth]{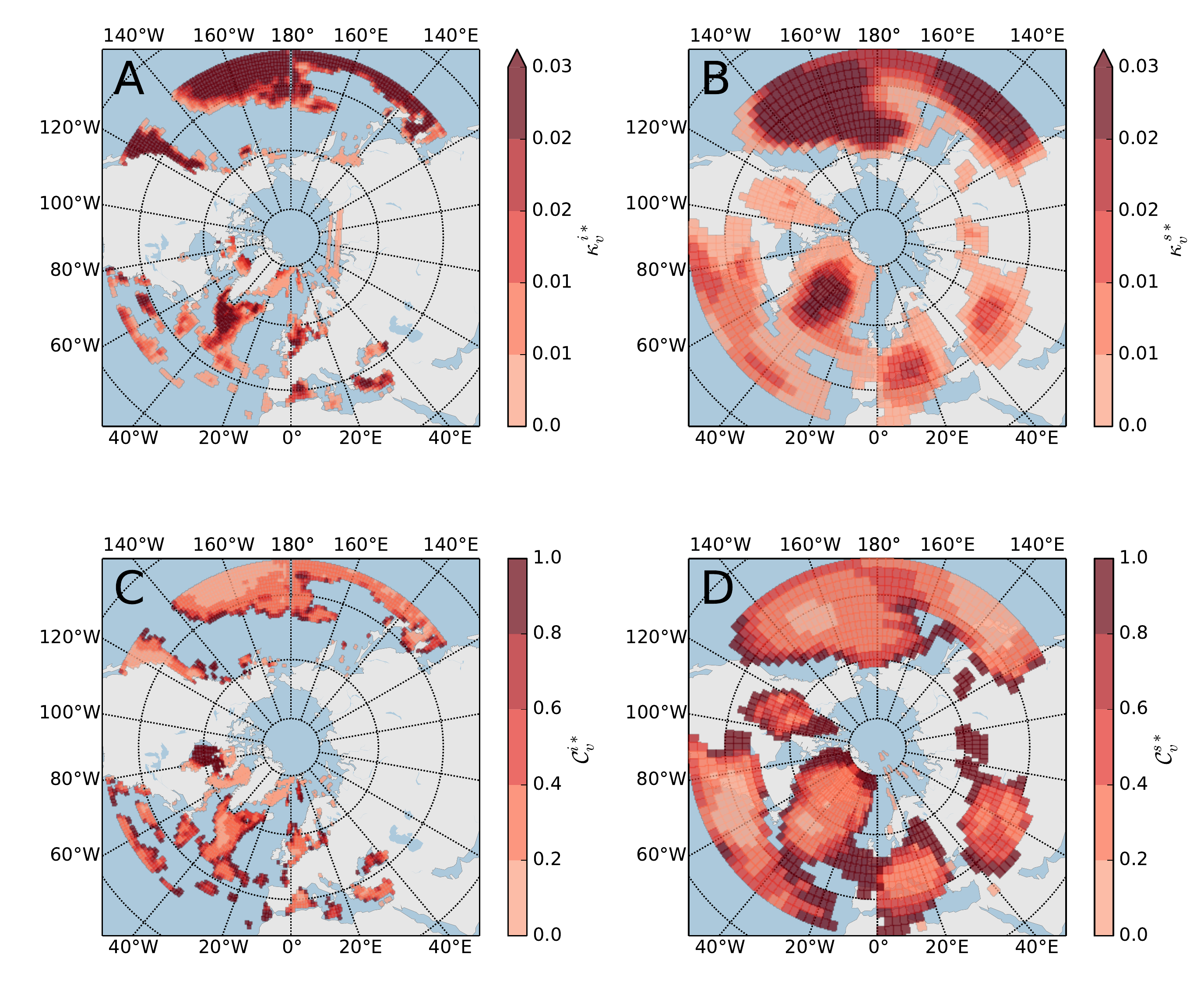} 
\caption{Coupled climate network analysis of ocean-atmosphere interactions northward of 30$^\circ$ N: N.s.i.\ cross-density (A) for nodes in the sea surface temperature field and (B) for nodes in the geopotential height field at 500 mbar. (C, D) The same for the n.s.i.\ local cross-clustering coefficient.}
\label{fig:ccn}
\end{figure}

Local interacting network measures allow for the detection of regions in one field that couple
with the other field and additionally provide a notion of the resulting
coupling strength and structure (Fig.~\ref{fig:ccn}). The \emph{n.s.i.\, cross-degree
density} 
\begin{align}
\kappa_v^{ij \ast} &= \frac{1}{W_j} \sum_{q \in V_j} w_q A_{vq}^+ \\
&= \frac{1}{W_j} k_v^{ij \ast}
\end{align}
measures the weighted share of nodes in another subnetwork $V_j$ that
each node $v \in V_i$ is connected with. It is obtained by normalizing the 
n.s.i.\, cross-degree $k_v^{ij \ast}$ (\texttt{InteractingNetworks.nsi\_cross\_degree}) by the sum
over all node weights in the opposite subnetwork $W_j = \sum_{q \in V_j} w_q$ such that its values range
between 0 and 1. Thus, high values indicate a strong localized coupling between
the fields or climate subnetworks. Figure~\ref{fig:ccn}A shows the n.s.i.\
cross-degree density $\kappa_v^{si \ast} \equiv \kappa_v^{i \ast}$ for nodes in the SST field (subnetwork index $s$). We find several localized
areas in the Atlantic as well as the Pacific that show strong coupling with the
HGT field (subnetwork index $i$). In contrast, the n.s.i\ cross-degree density $\kappa_v^{is \ast} \equiv \kappa_v^{s \ast}$ for nodes in the HGT
field shows large areas of pronounced coupling with the SST field (Fig.~\ref{fig:ccn}B).
It should be noted, however, that this measure by definition does not contain any information on the interactions within each of the fields. 

Additionally, the \emph{n.s.i.\ local cross-clustering coefficient}
\begin{align}
\mathcal{C}_v^{ij \ast} &= \frac{\sum_{p,q \in V_j} A^+_{vp} w_p A^+_{pq} w_q A^+_{qv}}{(k_v^{ij \ast})^2}
\end{align}
indicates whether two neighbors in subnetwork $V_j$ of a considered node $v \in V_i$ are also
mutually connected and, hence, measures the weighted share of triangular
structures between both subnetworks. It is computed using the method
\texttt{InteractingNetworks.nsi\_cross\_local\_clustering}. We note that
generally the n.s.i.\ cross-local clustering coefficient takes lower values for
nodes in the SST field ($\mathcal{C}_v^{si \ast} \equiv \mathcal{C}_v^{i \ast}$, Fig.~\ref{fig:ccn}C) then for nodes in the HGT field
($\mathcal{C}_v^{is \ast} \equiv \mathcal{C}_v^{s \ast}$, Fig.~\ref{fig:ccn}D) implying that nodes in the SST field couple
preferably with nodes in the HGT field, which are themselves dynamically
dissimilar and, hence, disconnected.  

In fact, we find that the ocean-to-atmosphere interaction in the Northern
hemisphere follows a hierarchical structure \citep{ravasz_hierarchical_2003},
meaning that larger areas of dynamically similar nodes in the SST field couple
with several dynamically dissimilar areas in the HGT field~\citep{Wiedermann2015}. This feature may be
attributed to large-scale ocean currents interacting with different parts of
the atmosphere along their respective directions of flow.

\section{Network-based time series analysis}
\label{sec:time_series}

Network-based methodologies provide valuable novel approaches to nonlinear time series analysis that have manifold applications ranging from studying the detailed geometrical structure of a dynamical system in phase space to detecting critical transitions or tipping points in observational time series \citep{Xu2008,Donner2011a}. While \emph{time series networks} can reflect the dynamical properties of time series obtained from a complex system in a smorgasbord of different ways, \texttt{pyunicorn} focusses on two complementary approaches: (i) \emph{Recurrence networks}~\citep{Marwan2009,Donner2010b}, an approach closely related to \emph{recurrence quantification analysis} of \emph{recurrence plots}, are random geometric graphs~\citep{Dall2002,Donges2012} representing proximity relationships (links) of state vectors (nodes) in phase space (Sect.~\ref{sec:recurrence}). (ii) \emph{Visibility graphs} encode visibility relations between data points in the one-dimensional time domain (Sect.~\ref{sec:visibility_graphs}; \citet{Lacasa2008,Donner2012visibility}). Hence, while recurrence networks allow to investigate geometric properties of the system such as the transitivity dimension~\citep{donner2011epjb}, visibility graphs can be applied to investigate purely temporal features such as long-range correlations~\citep{Lacasa2009} or time-reversal asymmetry~\citep{Donges2013}. Network-based time series analysis is demonstrated by discussing two use cases from paleoclimatology that aim at detecting regime shifts or tipping points in climate dynamics on longer time-scales.

\subsection{Recurrence analysis}
\label{sec:recurrence}

Recurrence is a fundamental property of many dynamical systems and is used by several approaches in order to investigate a system's dynamics.
A basic tool of nonlinear time series analysis is the binary \emph{recurrence matrix} \citep{marwan2007}
\begin{equation}
R_{pq}(\varepsilon) = \Theta(\varepsilon - ||\mathbf{x}(p) - \mathbf{x}(q)||), \label{eq:rp}
\end{equation}
where $\mathbf{x}(p)$ is a state vector at time $p = 1,\ldots,N$, $N$ the number of states,
$\Theta(\cdot)$ again the Heaviside function, and $\varepsilon$ the recurrence threshold.
A graphical representation, the \emph{recurrence plot}, provides visual, qualitative impressions
about the dynamics of even high-dimensional systems. Quantitative approaches based on this matrix have been developed
for studying different aspects of complex systems,
particularly based on univariate and multivariate time series data \citep{marwan2007}. Recurrence
quantification analysis (RQA) and recurrence network analysis (RNA) allow to
classify different dynamical regimes in time series and to detect regime
shifts, dynamical transitions or tipping points, among many other applications \citep{marwan2015b}.
Bivariate methods such as joint recurrence plots/networks, cross-recurrence
plots or inter-system recurrence networks can be used to investigate the coupling
structure between two dynamical systems based on their time series, including methods
to detect the directionality of coupling \citep{romano2007estimation,zou2011inferring,Feldhoff2012}. Recurrence analysis is applicable to
general time series data from many fields such as climatology, medicine, neuroscience or economics \citep{marwan2008epjst}. These applications range from using recurrence analysis as a classifier for monitoring health states in medicine and engineering~\citep{ramirez2013} to detecting continental-scale nonlinear regime shifts in the Asian monsoon system during the Holocene~\citep{Donges2015b}.

\subsubsection{Recurrence quantification analysis}
\label{sec:rqa}

Recurrence of a dynamical system is usually studied in phase space. A standard approach is
to use time-delay embedding for reconstructing the appropriate phase space representation
from a measured time series \citep{Packard1980}.

\begin{figure}[tbp]
\begin{center}
\includegraphics[width=\columnwidth]{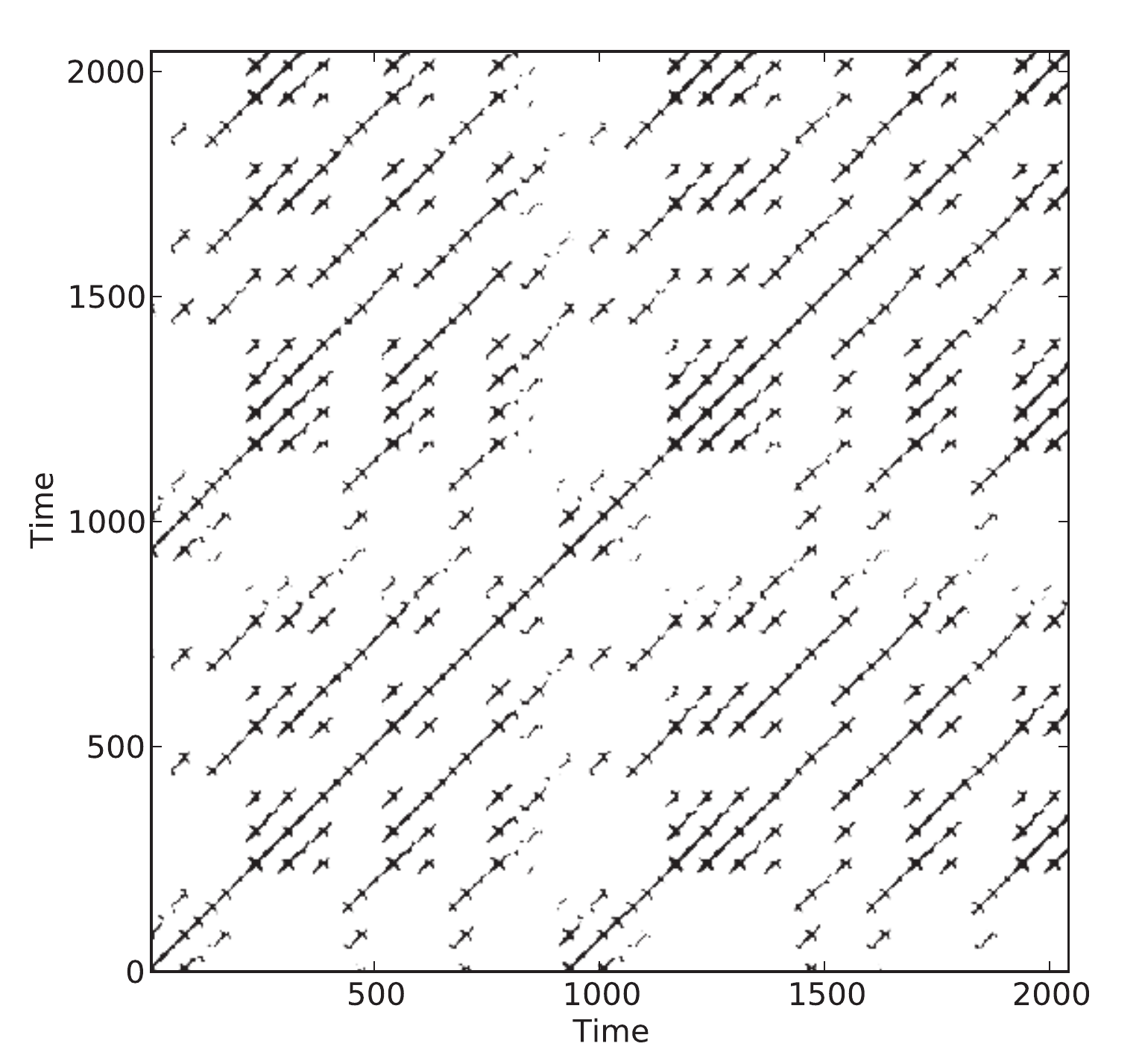}
\caption{Recurrence plot of a time series generated by the Lorenz '63 system in a chaotic regime~\citep{Lorenz1963}.}
\label{fig:rp_lorenz}
\end{center}
\end{figure}

The class \texttt{timeseries.RecurrencePlot} can be used to generate recurrence plots and perform
recurrence quantification analysis. The parameters \texttt{dim} and \texttt{tau} can be used
to set the parameters of the time-delay embedding, whereas \texttt{threshold} or \texttt{recurrence\_rate}
as well as \texttt{metric} configure the recurrence criteria (i.e. used recurrence norm and threshold $\varepsilon$).
The method \texttt{RecurrencePlot.embedding} can be used to get the reconstructed phase space vectors resulting from the specified embedding parameters. For example, a recurrence plot can be computed from a given scalar time series $x(t)$ using the following code:
\begin{verbatim}
rp = RecurrencePlot(x, recurrence_rate=0.05, 
dim=3, tau=30)
\end{verbatim}
The recurrence matrix $\mathbf{R}$ (Eq.~\ref{eq:rp}) can be extracted as the property \texttt{RecurrencePlot.R} and can be plotted (Fig.~\ref{fig:rp_lorenz}), e.g. using \texttt{matplotlib}, or used for subsequent analysis.

The simplest quantifier in recurrence analysis is the probability that any recurrence will occur, i.e. the
fraction positive entries in $\mathbf{R}$, called {\it recurrence rate} (\texttt{RecurrencePlot.recurrence\_rate}). However, information about dynamical properties of the system is represented by the diagonal and vertical lines apparent in the recurrence plot. The line length distributions are, thus, the foundation for statistical, quantitative analysis of the recurrence matrix $\mathbf{R}$ (Eq.~(\ref{eq:rp})), called {\it recurrence quantification analysis} \citep{marwan2007}. Moreover, the empty vertical spaces in $\mathbf{R}$, apparent as white vertical lines in the recurrence plot, correspond to recurrence times.
Several measures of complexity using these line distributions (diagonal, vertical, and white lines) are available as methods in the \texttt{RecurrencePlot} class, e.g. maximal diagonal line length (\texttt{max\_diaglength}), determinism (\texttt{determinism}), laminarity (\texttt{laminarity}), diagonal line entropy (\texttt{diag\_entropy}) or mean recurrence time (\texttt{mean\_recurrence\_time}). The distributions of diagonal and vertical lines (\texttt{diagline\_dist} and \texttt{vertline\_dist}) can be useful for further quantifications, e.g. by looking at the scaling behavior, which is related to the $K_2$ entropy \citep{marwan2007}. Resampled instances of both types of line distributions can be generated using the methods \texttt{resample\_diagline\_dist} and \texttt{resample\_vertline\_dist} for estimating confidence bounds for RQA measures following the permutation-based method proposed by~\citep{schinkel2009confidence}.

\texttt{pyunicorn} furthermore supports multivariate extensions of RQA such as joint (\texttt{timeseries.JointRecurrencePlot}) and cross recurrence plots (\texttt{timeseries.CrossRecurrencePlot}) that both inherit from the \texttt{RecurrencePlot} class.

\subsubsection{Recurrence network analysis}
\label{sec:recurrence_networks}

\begin{figure}[tbp]
\begin{center}
\includegraphics[width=\columnwidth]{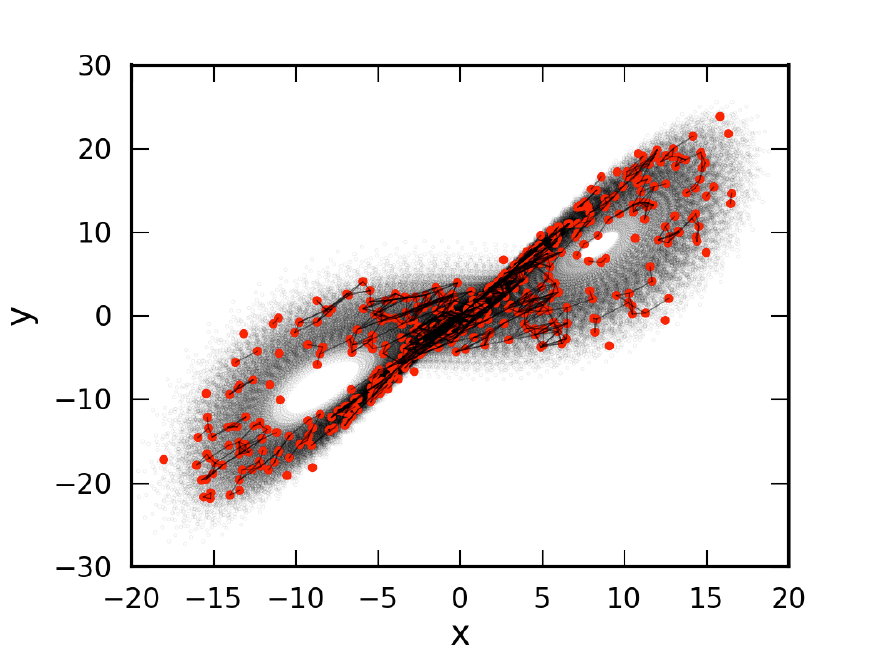}
\caption{Recurrence network as a random geometric graph generated from $N=750$ state vectors (red filled circles) drawn from the invariant density of the Lorenz '63 system in a chaotic regime~\citep{Lorenz1963}. In this example, state vectors (nodes) with a distance smaller than $\varepsilon=2$ according to the supremum norm are connected by a link.}
\label{fig:rn_lorenz}
\end{center}
\end{figure}

The striking similarity of the binary square recurrence matrix (Eq.~\ref{eq:rp}) with the adjacency matrix (Eq.~\ref{eq:adjacency}) of an unweighted
and undirected network has lead to a complementary kind of recurrence analysis by measures from
complex network studies \citep{Marwan2009,Donner2010b}. More formally, the \emph{recurrence networks} (Fig.~\ref{fig:rn_lorenz}) defined in this way by their adjacency matrix
\begin{align}
A_{pq}(\varepsilon) = R_{pq}(\varepsilon) - \delta_{pq},
\end{align}
where $\delta_{pq}$ is Kronecker's delta introduced to avoid self-loops in the networks, can be understood as random geometric graphs (Fig.~\ref{fig:rn_lorenz}) that capture rich information on the geometrical structure of a dynamical system's invariant density in phase space~\citep{Dall2002,Donges2012}. The nodes in a recurrence network represent state vectors and the links indicate proximity relationships between them. Recurrence networks are represented by the \texttt{timeseries.RecurrenceNetwork} class inheriting from \texttt{Network} and \texttt{RecurrencePlot} (Fig.~\ref{fig:uml}A).

This approach opens up the wealth of measures, models and algorithms from complex network theory for time series analysis. Network measures such as average path length (\texttt{Network.average\_path\_length}), global clustering coefficient (\texttt{Network.global\_clustering}), transitivity (\texttt{Network.transitivity}) or assortativity (\texttt{Network.assortativity}) characterize the geometrical properties of the dynamical system trajectories in phase space and can be used to differentiate between dynamical regimes,e.g. periodic and chaotic~\citep{Marwan2009,Donner2010b,zou2010}. In particular, the \emph{transitivity}
\begin{align}
\mathcal{T}(\varepsilon) = \frac{\sum_{v,p,q \in V} A_{vp}(\varepsilon) A_{pq}(\varepsilon) A_{vq}(\varepsilon)}{\sum_{v,p,q \in V} A_{vp}(\varepsilon) A_{vq}(\varepsilon)}
\end{align}
is appropriate because it can be linked to the geometry of the phase space trajectory \citep{donner2011epjb,Donges2012b}. Specifically, it can be logarithmically transformed to yield the \emph{single-scale transitivity dimension}
\begin{align}
D_{\mathcal{T}}(\varepsilon) = \frac{\log \mathcal{T}(\varepsilon)}{\log(3/4)},
\end{align}
a global dimension-like measure of the geometric organization of the available set of state vectors in phase space (\texttt{RecurrenceNetwork.transitivity\_dim\_single\_scale}). Analogously, a transformed local clustering coefficient yields a local dimension-like measure that is defined on every node or state vector (\texttt{RecurrenceNetwork.local\_clustering\_dim\_single\_ scale}).

Analogously to multivariate RQA (Sect.~\ref{sec:rqa}), multivariate extensions of recurrence network analysis have been applied to investigate directions of coupling between dynamical systems~\citep{Feldhoff2012} and complex synchronization scenarios including generalized synchronization~\citep{Feldhoff2013}. The corresponding methodologies are represented by the classes \texttt{timeseries.InterSystemRecurrenceNetwork} and \texttt{timeseries.JointRecurrenceNetwork}, respectively.

\subsubsection{Use case: identification of transitions in paleoclimate variability}

A relevant application of recurrence analysis is the detection of dynamical transitions in complex systems captured by model or observational time series~\citep{Marwan2009,Donges2011b}. Detecting such dynamical transitions, regime shifts or tipping points~\citep{Lenton2008,rockstrom2009} is of great interest in studying past climate variability to gain a deeper understanding of the Earth's climate system also on geological time scales~\citep{Donges2011c,Donges2015b}. In the following we discuss a typical example from paleoclimate research following \citet{marwan2015b} that focusses on investigating interactions between sea-surface temperature (SST) and the dynamics of specific climate subsystems, such as the Asian monsoon system or the Atlantic thermohaline circulation, as well as regime shifts therein.

\begin{figure}[tbp]
\begin{center}
\includegraphics[width=\columnwidth]{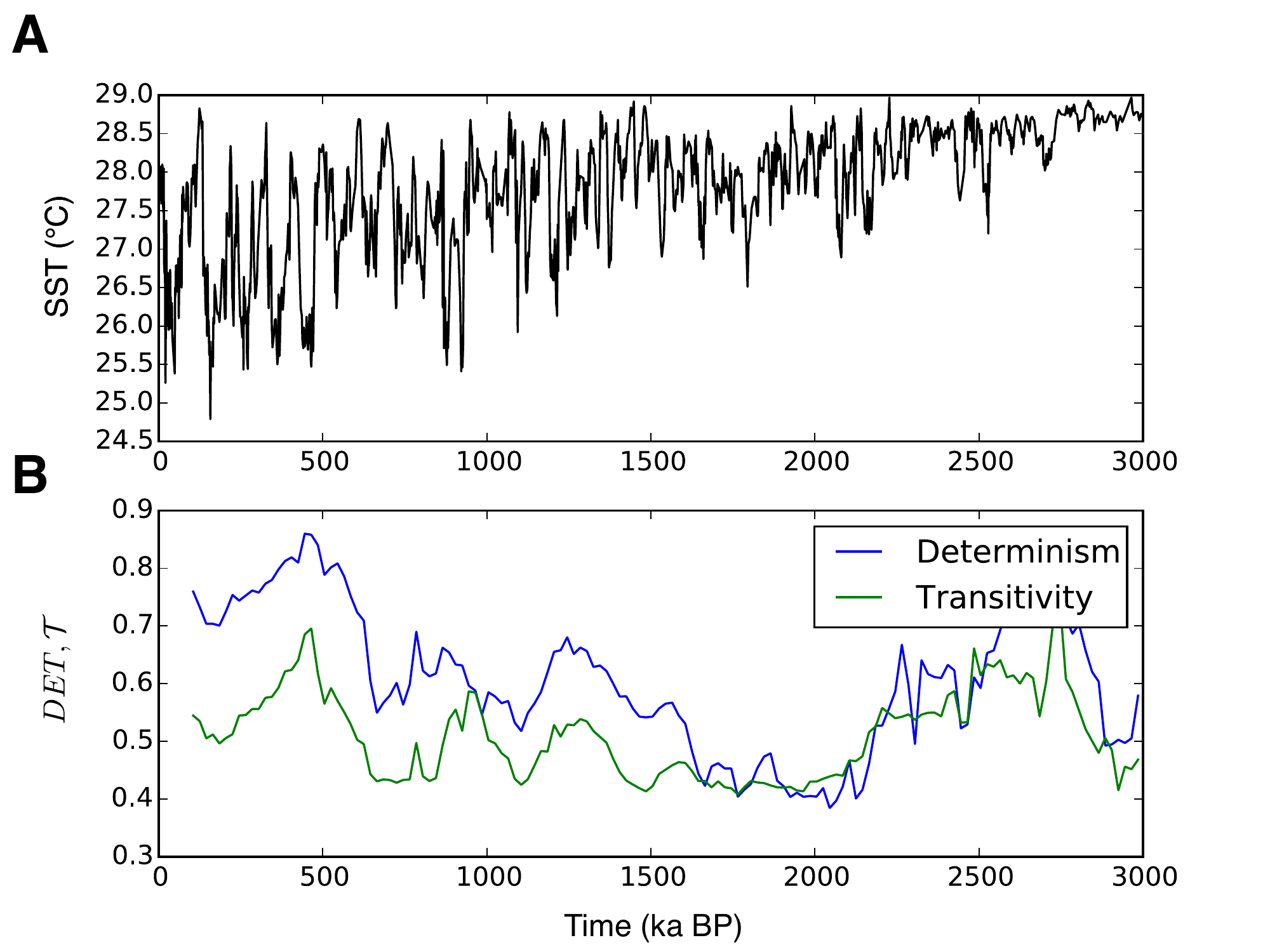}
\caption{Studying dynamical transitions in paleoclimate data using recurrence analysis: (A) Alkenone paleothermometry based SST estimates for the South China Sea and (B) corresponding determinism DET and transitivity $\mathcal{T}$.
}
\label{odp1143_transit}
\end{center}
\end{figure}

Diverse types of geological archives are used in paleoclimatology to reconstruct and study climate
conditions of the past, such as lake \citep{marwan2003climdyn} and marine sediments \citep{herbert2010,Donges2011b,Donges2011c} or
speleothems \citep{kennett2012,Donges2015b}. Alkenone remnants in the organic fraction of
marine sediments, produced by phytoplankton, can be used to reconstruct SSTs of the past (alkenone paleothermometry), which allows to investigate  past oceanic temperature variability~\citep{herbert2001,li2011epsl}.
In this use case, we analyze an SST reconstruction for the South China Sea covering the past 3~Ma 
that is derived from alkenone paleothermometry of the Ocean Drilling Programme (ODP) site 1143 drill core~\citep{li2011epsl}
(Fig.~\ref{odp1143_transit}A). The South China Sea is strongly connected to the East Asian Monsoon (EAM) system
encompassing a winter monsoon season with strong winds and a summer monsoon season with particularly high precipitation.

We generate \texttt{RecurrenceNetwork} objects and compute the measures determinism DET (\texttt{RecurrenceNetwork.determinism}) and transitivity $\mathcal{T}$ (\texttt{RecurrenceNetwork.transitivity}) for sliding windows of length 410\,ka 
(containing a varying number of data points due to the time series' irregular sampling) 
and a step size of 20\,ka (Fig.~\ref{odp1143_transit}B).
For reconstructing the phase space by time-delay embedding~\citep{Packard1980}, we select an embedding dimension of $6$ (as 
suggested by the false nearest neighbors criterion~\citep{kennel92}).
The selection of the time-delay parameter is guided by the auto-correlation function. As a result, it is approximated as 20\,ka for
all time windows based on the median sampling time within each window. 
The recurrence threshold is chosen to preserve a prescribed recurrence rate of 7.5\% \citep{marwan2007,Donges2011c}.

During the past 3~Ma, several major and many smaller climate changes occurred on regional, but
also global scales. Particularly pronounced climate shifts have been related to Milankovich cycles \citep{haug1998,medinaelizalde2005,an2014}
and major changes in ocean circulation patterns \citep{karas2009}. 
Following a transition towards obliquity-driven climate variability
with a 41\,ka period around 3.0\,Ma BP (before present), a long period of globally warm climate ended and Northern hemisphere glaciations
started after 2.8--2.7\,Ma BP~\citep{haug1998,herbert2010,an2014}. This transition is revealed by a 
significant increase of DET and $\mathcal{T}$ between 2.8 and 2.2\,Ma signifying an increased regularity of SST dynamics during this period. From detailed studies of loess sediments it is known that around 1.25\,Ma BP, the intensity of the EAM winter monsoon season started to be strongly coupled to global ice-volume change \citep{an2014}. Around this time also marking the beginning of a transition phase towards glacial-interglacial cycles of 100\,ka period (eccentricity-dominated period of the Milankovich cycles), DET and $\mathcal{T}$ increase markedly again.
Dominance of the 100\,ka period was well established after 0.6\,Ma and is clearly indicated by increased DET and $\mathcal{T}$ values between 0.6 and 0.2\,Ma BP \citep{sun2010}.
It is also known from loess sediments that the EAM summer monsoon weakened between 2.0 and 1.5\,Ma BP and
around 0.7\,Ma BP. During these periods, DET and $\mathcal{T}$ assume lower values and, hence, more irregular SST variability, than during the previously discussed epochs. 

In this way, recurrence analysis by means of the measures DET and $\mathcal{T}$ confirms earlier findings of strong links between the EAM and Milankovich cycles. Furthermore, the analysis of recurrence properties allows for deeper insights, such as that dominant Milankovich cycles and periods of major climate transitions from one to another regime go along with increased and reduced regularity in the (regional) climate dynamics in the East Asian Monsoon system (as reflected by the South China Sea SST and for the considered time scales).

\subsection{Visibility graphs}
\label{sec:visibility_graphs}


\emph{Visibility graph} (VG) methods represent an alternative approach for transforming time series into complex networks \citep{Lacasa2008}, which draws upon analogies between height profiles in physical space and the profile of a time series graph. Originally utilized in fields like architecture and robot motion planning, VGs are based on the existence or non-existence of lines of sight between well-defined objects. 

\subsubsection{Theory of time series visibility graphs}

In a time series context, these objects are the sampling points of a (univariate) time series graph, which are uniquely characterized by pairs $(t_v,x_v)$ with $x_v=x(t_v)$. From a practical perspective, we can identify each node $v$ of a \emph{standard visibility graph} with a given time point $t_v$. For $t_v<t_p$ (and, hence, $v<p$) a link between the nodes $v$ and $p$ exists iff
\begin{equation}
x_q < x_v + \frac{x_p-x_v}{t_p-t_v}(t_q-t_v) \ \forall \ v<q<p.
\label{eq:vg}
\end{equation}
\noindent
Put differently, the topological properties of VGs are closely related to the roughness of the underlying time series profile.

As a notable algorithmic variant, \emph{horizontal visibility graphs} (HVGs) facilitate analytical investigations of the graph profile \citep{Luque2009PRE}. In this case, Eq.~(\ref{eq:vg}) is replaced by the simpler condition
\begin{equation}
x_q < \min\{x_v,x_p\} \ \forall \ v<q<p.
\label{eq:hvg}
\end{equation}
\noindent
One easily convinces oneself that the latter condition is more restrictive, and that the link set of a HVG is a subset of that of the standard VG. In turn, the classical VG is invariant under arbitrary affine transformations of a time series, whereas the HVG retains this invariance only for uniform translations and rescaling of the original data.

In \texttt{pyunicorn}, VG and HVG can be generated from any time series object via the \texttt{timeseries.VisibilityGraph} class. The decision on which of the two variants is used is mediated via the Boolean parameter \texttt{horizontal}, which can be set to \texttt{False} (standard VG, default value) or \texttt{True} (HVG). Additional information on the timings of individual observations as well as missing values can be provided via supplementary optional parameters.

The distinctively different construction mechanism of (H)VGs in comparison to recurrence networks (Sect.~\ref{sec:recurrence_networks}) implies a completely different interpretation of the resulting networks. First, recurrence networks are equally applicable to uni- and multivariate time series, whereas VGs are tailored for applications to univariate records. So far, there is no way to unambiguously generalize the VG concept to bi- or multivariate time series. 

Second, both types of time series networks are spatial networks (Sect.~\ref{sec:spatial_networks}) in an abstract sense - recurrence networks' nodes being characterized by the positions of the associated state vectors in the dynamical system's phase space, and those of VGs being fixed along the time axis representing some abstract one-dimensional space. This fact implies strong restrictions to the resulting graph properties \citep{Donner2012visibility}, including potentially severe biases of (especially path-based) network properties due to the absence of information prior to the first, as well as after the last sampling point of the time series. To this end, simplified functionalities correcting for such effects have been implemented for the case of degree and closeness centrality measures (\texttt{boundary\_corrected\_degree} and \texttt{boundary\_corrected\_closeness}, respectively). An improved treatment is planned for future releases of \texttt{pyunicorn}, but still requires further analytical understanding of the VG properties.

Third, while power-law degree distributions of recurrence networks are related with specific singularities of the probability density of states \citep{Zou2012EPL}, they are a widely observable feature of VGs that arises from the presence of long-range correlations \citep{Lacasa2009,Ni2009PLA}. Specifically, under the assumption of a fractional Brownian motion or related stochastic process, the characteristic scaling exponent of the degree distribution of a VG can be directly related with the Hurst exponent of the series, whereas there are no similar unique relationships for recurrence networks \citep{Zou2012EPL}. However, given the multiplicity of existing estimators of the latter quantity, the practical advantage of using VGs and HVGs for this purpose has not yet been convincingly demonstrated.

Finally, given the direct association between nodes and time points, VGs and HVGs provide a means to discriminate the statistical properties of a graph when looking forward and backward in time, respectively, from a given observation point. This gives birth to a new class of time series irreversibility tests \citep{Lacasa2012EPJB,Donges2013}, which allow to evaluate the null hypothesis of linearity since irreversibility is a common hallmark of nonlinear dynamics \citep{Theiler1992PhysD}. Specifically, local (node-wise) graph properties can be decomposed into contributions from either past or future observations (\emph{retarded} (backward) vs. \emph{advanced} (forward)), as well as such combining information from past and future (\emph{trans}). Corresponding features are implemented in \texttt{pyunicorn} for a variety of network properties, including degree (\texttt{retarded\_degree} vs. \texttt{advanced\_degree}), local clustering coefficient (\texttt{retarded/advanced\_local\_clustering}), closeness (\texttt{retarded/advanced\_closeness}) and betweenness (\texttt{retarded/advanced/trans\_betweenness}).


\subsubsection{Use case: time irreversibility in glacial-deglacial dynamics}

\begin{figure*}[tbp]
   \centering
   \includegraphics[width=0.8\textwidth]{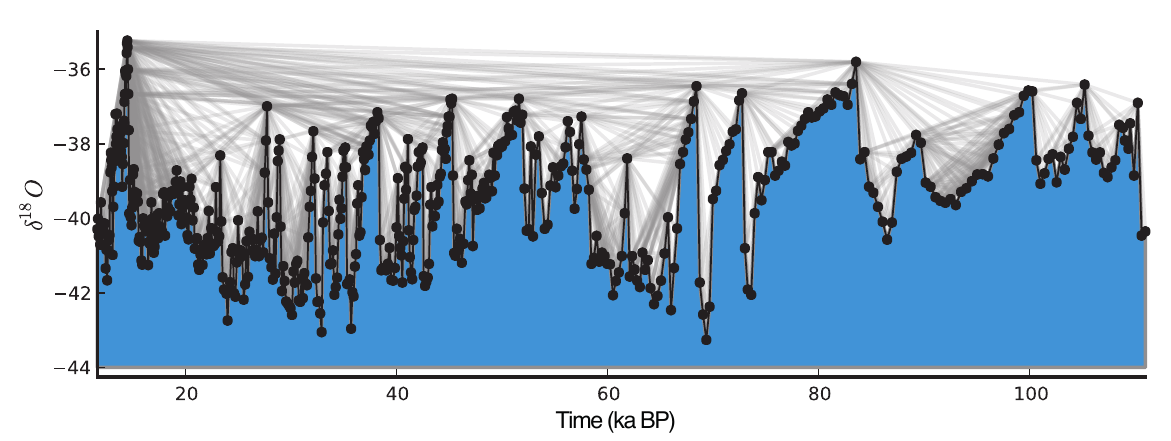}
   \caption{Visibility graph of the GISP2 ice core $\delta^{18} O$ record (2m resolution) from Greenland during the last glacial~\citep{Grootes1997}. Black dots indicate observations (nodes). Links are added between mutually visible observations (gray lines).}
   \label{fig:gisp2_plot}
\end{figure*}

VG-based tests for time series irreversibility have been successfully applied to discriminate between different dynamical regimes in EEG recordings from healthy and epileptic patients~\citep{Donges2013}. It is instructive to also apply this methodology to study regime shifts in paleoclimate dynamics~\citep{Schleussner2015}, particularly for investigating in depth the important differences between interglacial and glacial climate variability and the nature of the transitions between those regimes~\citep{Donges2012b}. As an example, we study the 2\,m resolution $\delta^{18} O$ isotope record from the GISP2 ice core from Greenland~\citep{Grootes1997} convering both the Holocene and the last glacial period (Fig.~\ref{fig:gisp2_plot}). The boundary between the last glacial and the Holocene is defined by the Younger Dryas-Preboreal transition at about 11,650 years BP~\citep{Grootes1997}. 

We apply the test for time irreversibility proposed in~\citet{Donges2013}, which is based on comparing the distributions of \emph{time-directed degree} (\texttt{VisibilityGraph.retarded/advanced\_degree}) 
\begin{align}
k_v^r &= \sum_{p \in V: p < v} A_{vp},\\
k_v^a &= \sum_{p \in V: p>v} A_{vp}
\end{align}
and \emph{time-directed local clustering coefficient} (\texttt{VisibilityGraph. retarded/advanced\_local\_clustering})
\begin{align}
\mathcal{C}_v^r &= {k_v^r \choose 2}^{-1} \sum_{p,q \in V: p,q < v} A_{vp} A_{pq} A_{qv},\\
\mathcal{C}_v^a &= {k_v^a \choose 2}^{-1} \sum_{p,q \in V: p,q > v} A_{vp} A_{pq} A_{qv}.
\end{align}
using the Kolmogorov-Smirnov test. To ensure the robustness of the results, a $k$-leave-out cross-validation is conducted by generating ensembles of $100$ realizations for each record by bootstrapping 80\% of the data points without changing their time ordering and, subsequently, performing the test to each ensemble member. Following this protocol for the Holocene and the last glacial period time series segments separately, we find that the null hypothesis of reversibility can be quite safely rejected for the last glacial, whereas this is not the case for the Holocene (Tab.~\ref{tab:gisp_vg_reversibility_test}). Since the conditions for the deposition of ice and snow at the drilling site are argued to have remained more or less constant throughout the entire time span covered by the record, these results point at a strong signature of nonlinear climate dynamics during the last glacial, which is not detected for the Holocene, consistently with results for Holocene speleothem records~\citep{Donges2012b}. One visually directly accessible piece of evidence for the indicated irreversible dynamics during the last glacial period are the frequently occurring Dansgaard-Oeschger events that are characterized by rapid warming (change towards more positive $\delta^{18} O$ values) and subsequent slower cooling (change towards more negative $\delta^{18} O$ values) of Greenland climate (Fig.~\ref{fig:gisp2_plot}). Note that the sampling times increase with age for the GISP2 as for all ice core proxy records (in GISP2, the average sampling time is $14.2 \pm 6.3$\,a for the Holocene, while it is $175.8 \pm 119.6$\,a during the last glacial period), because the ice column is compressed by ice and snow deposited on top of it over the years. It remains to be tested in future research whether this nonstationarity has a strong impact on the results of the VG test for time irreversibility.

\begin{table}[tbp]
\caption{Results of the visibility graph-based tests for time irreversibility of the GISP2 ice core $\delta^{18} O$ record from Greenland~\citep{Grootes1997}. The number of data points in each time series segment is denoted by $M$, while $q_k$ and $q_\mathcal{C}$ give the rate of rejection for the null hypothesis of time reversibility (at a 95\% significance level) during a $k$-leave-out cross-validation for degree and clustering-based tests, respectively.}
\begin{center}
\begin{tabular}{llrrrrr}
\hline
Period & Time span (ka BP) & M & $q_k$ & $q_\mathcal{C}$ \\
\hline
Last glacial & 110.98 -- 11.65 & $566$ & $0.93$ & $0.82$ \\
Holocene & 11.65 -- today & $824$ & $0.00$ & $0.00$ \\
\hline
\end{tabular}
\end{center}
\label{tab:gisp_vg_reversibility_test}
\end{table}%

\section{Surrogate time series}
\label{sec:surrogates}

As mentioned in several contexts above and analogously to random network models (see, e.g. Sect.~\ref{sec:spatial_networks} on random surrogate models for spatial networks and Sect.~\ref{sec:interacting_networks} on surrogate models for networks of networks), \emph{surrogate time series} are a useful methodology for testing the statistical significance of observed time series properties such as those derived from functional networks or network-based time series analysis based on various null hypotheses~\citep{Schreiber2000}. The idea behind this approach is to generate surrogate time series that conserve certain properties of observed time series such as the amplitude distribution or auto-correlation function, but are random otherwise. \texttt{pyunicorn} can be used to generate several commonly used types of time series surrogates (\texttt{timeseries.Surrogates} class): \emph{white noise surrogates}, \emph{Fourier surrogates}, \emph{amplitude adjusted Fourier transform surrogates}~\citep{Schreiber2000} or \emph{twin surrogates}~\citep{Thiel2006twin}. White noise surrogates correspond to randomly shuffled copies of a time series $x(t)$ and, hence, retain only the amplitude distribution of the original data (\texttt{Surrogates.white\_noise\_surrogates}). In contrast, Fourier surrogates are generated by randomizing the phase components of $x(t)$ in Fourier space and conserve the time series' power spectrum (and thus also its linear auto-correlation function via the Wiener-Chintschin theorem), but not the amplitude distribution (\texttt{Surrogates.correlated\_noise\_surrogates}). This potential drawback is addressed by amplitude adjusted Fourier transform (AAFT) surrogates that only approximate the power spectrum but conserve the amplitude distribution (\texttt{Surrogates.AAFT\_surrogates}). An even closer match is provided by iteratively refined AAFT surrogates (\texttt{Surrogates.refined\_AAFT\_surrogates}). Finally, twin surrogates approximate the recurrence structure of the original time series and, hence, have the potential to additionally conserve certain nonlinear properties of $x(t)$ such as the auto-mutual information function (\texttt{Surrogates.twin\_surrogates}). Note that the surrogate types supported by \texttt{pyunicorn} only conserve certain properties of single time series, whereas multivariate surrogate methods~\citep{Schreiber2000} conserving properties of pairs of time series such as the linear cross-correlation function are currently not implemented.

When computing functional networks, surrogate time series can be used to include links in the network based on a fixed significance level instead of a fixed threshold or link density in terms of the considered similarity measure from coupling analysis (see, e.g. \citet{palus2011}). In network-based time series analysis, surrogate time series provide a means to test the statistical significance of results obtained for the data at hand based on a hierarchy of null hypotheses, for example, when searching for dynamical transitions and regime shifts~\citep{Donges2011b,Donges2011c} or when studying the directionality of coupling or synchronization between time series~\citep{Feldhoff2012,Feldhoff2013}.

\section{Conclusions and perspectives}
\label{sec:conclusions}

In this article, we have described the \texttt{pyunicorn} software package, which facilitates the study of various types of complex networks as well as a detailed investigation of time series data using modern methods of functional network and network-based nonlinear time series analysis. \texttt{pyunicorn} is written in the programming language Python and, hence, is conveniently applicable to research domains in science and society as different as neuroscience, infrastructure and climatology. Most computationally demanding algorithms are implemented in fast compiled languages on sparse data structures, allowing the performant analysis of large networks and time series data sets. The software's modular and object-oriented architecture enables the flexible and parsimonious combination of data structures, methods and algorithms from different fields. For example, combining complex network theory (\texttt{core.Network} class) and recurrence plots (\texttt{timeseries.RecurrencePlot} class) yields recurrence network analysis (\texttt{timeseries.RecurrenceNetwork} class), a versatile framework that opens new perspectives for nonlinear time series analysis and the study of complex dynamical systems in phase space. Another example are climate networks (\texttt{climate.ClimateNetwork} class), an approach bringing together ideas and concepts from complex network theory and classical eigen analysis of climatological data (e.g. empirical orthogonal function analysis)~\citep{Donges2015}.

Along these lines, \texttt{pyunicorn} has the potential to facilitate future methodological developments in the fields of network theory, time series analysis and complex systems science by synthesizing existing elements and by adding new methods and classes that interact with or build upon preexisting ones. Nonetheless, we urge users of the software to ensure that such developments are theoretically well-founded and explicable as well as motivated by well-posed and relevant research questions to produce high-quality research.

\acknowledgements

This work has been financially supported by the Leibniz association (project ECONS), the German National Academic Foundation (Studienstiftung des deutschen Volkes), the Federal Ministry for Education and Research (BMBF) via the Potsdam Research Cluster for Georisk Analysis, Environmental Change and Sustainability (PROGRESS), the BMBF Young Investigators Group CoSy-CC2 (grant no. 01LN1306A), BMBF project GLUES, the Stordalen Foundation (Norway), IRTG~1740~(DFG) and Marie-Curie ITN LINC (P7-PEOPLE-2011-ITN, grant No.\,289447). We thank Kira Rehfeld and Nora Molkenthin for helpful discussions. Hanna C.H. Schultz, Alraune Zech, Jan H. Feldhoff, Aljoscha Rheinwalt, Hannes Kutza, Alexander Radebach, Alexej Gluschkow, Paul Schultz, Stefan Schinkel, and Wolfram Barfuss are acknowledged for contributing to the development of \texttt{pyunicorn} at different stages. We thank all those people who have helped improving the software by testing, using, and commenting on it. \texttt{pyunicorn} is available at \texttt{https://github.com/pik-copan/pyunicorn} as a part of PIK's TOCSY toolbox. The distribution includes an extensive online documentation system with the detailed API documentation also being available in the PDF format~\citep{Supplement}. The software description in this article as well as in the supplemental material~\citep{Supplement} are based on the \texttt{pyunicorn} release version 0.5.0.


%

\end{document}